\documentclass[prd,aps,10pt,showkeys,nofootinbib,showpacs,twocolumn]{revtex4-1}

\usepackage{amsmath,amssymb,amsfonts,amsthm}
\usepackage{amsbsy} 
\usepackage{epsfig}
\usepackage{latexsym}
\input amssym.def
\input amssym.tex

\usepackage{hyperref}

\newcommand{\oarX}[1]{\href{http://arxiv.org/abs/#1}{{\ttfamily #1}}}
\newcommand{\arX}[1]{\href{http://arxiv.org/abs/#1}{{\ttfamily arXiv:#1}}}

\def\barr{\begin{array}}
\def\earr{\end{array}}
\def\half{\frac{1}{2}}
\def\ben{\begin{equation}}
\def\een{\end{equation}}
\def\bs{\begin{subequations}}
\def\es{\end{subequations}}
\def\bena{\begin{eqnarray}}
\def\eena{\end{eqnarray}}

\def\const{{\rm constant}}
\def\bR{\mathbb{R}}

\def\im{{\rm i}}

\def\be{\begin{equation}}
\def\ee{\end{equation}}

\def\bes{\begin{eqnarray}}
\def\ees{\end{eqnarray}}

\begin{document}

\title{Quantum propagation across cosmological singularities}

\author{Steffen Gielen}
\affiliation{Theoretical Physics, Blackett Laboratory, Imperial College London, London SW7 2AZ, United Kingdom}
\affiliation{Perimeter Institute for Theoretical Physics, Waterloo, Ontario N2L 2Y5, Canada}
\affiliation{Canadian Institute for Theoretical Astrophysics (CITA), 60 St George Street, Toronto, Ontario M5S 3H8, Canada}
\author{Neil Turok}
\affiliation{Perimeter Institute for Theoretical Physics, Waterloo, Ontario N2L 2Y5, Canada}

\date{\today}

%%%%%%%%%%%%%%%%%%%%%%%%%%%%%%%%%%%%%%%%%%%%%%%%%%%%%%%

\begin{abstract}
The initial singularity is the most troubling feature of the standard cosmology, which quantum effects are hoped to resolve. In this paper, we study quantum cosmology with conformal (Weyl invariant) matter. We show that it is natural to extend the scale factor to negative values, allowing a large, collapsing universe to evolve across a quantum ``bounce" into an expanding universe like ours. We compute the Feynman propagator for Friedmann-Robertson-Walker backgrounds exactly, identifying curious pathologies in the case of curved (open or closed) universes. We then include anisotropies, fixing the operator ordering of the quantum Hamiltonian by imposing  covariance under field redefinitions and again finding exact solutions. We show how complex classical solutions allow one to circumvent the singularity while maintaining the validity of the semiclassical approximation. The simplest isotropic universes sit on a critical boundary, beyond which there is qualitatively different behavior, with potential for instability. Additional scalars improve the theory's stability. 
Finally, we study the semiclassical propagation of inhomogeneous perturbations about the flat, isotropic case, at linear and nonlinear order, showing that, at least at this level, there is {\it no} particle production across the bounce. These results form the basis for a promising new approach to quantum cosmology and the resolution of the big bang singularity. 
\end{abstract}


\maketitle

\section{Introduction}

On the largest scales we can observe, our Universe has a remarkably simple structure. It is homogeneous, isotropic, and spatially flat to very high accuracy. Furthermore, the primordial curvature fluctuations which seeded the formation of structure apparently took an extremely minimal form: a statistically homogeneous, Gaussian-distributed pattern of very small-amplitude curvature perturbations, with an almost perfectly scale-invariant power spectrum. While inflationary models are capable of fitting the data, it is nonetheless tempting to look for a simpler and more fundamental explanation. The early Universe was dominated by radiation, a form of matter without an intrinsic scale. In fact, it is believed that any well-defined quantum field theory must possess a UV fixed point, signifying conformal invariance at high energies. These lines of argument encourage us to investigate a minimal early universe cosmology, namely a quantum universe filled with conformally invariant matter~\cite{letter}. We start by studying the quantum propagation of homogeneous background universes, uncovering a number of surprising features. Then, we include inhomogeneous perturbations, treated semiclassically and perturbatively at both linear and nonlinear order. We do not, in this paper, propose a realistic scenario. Nor do we proceed far enough to study the effects of renormalization and the running of couplings, although these are no doubt important. Our focus is on the technical calculation of the causal (Feynman) propagator in some specific (and quasirealistic) cosmologies. We also postpone a discussion of the important question of the {\it interpretation} of the propagator and its use to compute probabilities to future work. Nevertheless, we believe our findings are instructive and form a useful starting point for such investigations.  

The simplest example of conformal matter is  a perfect fluid of radiation. In the context of cosmology, this is extremely well motivated since the early Universe was, we believe, radiation dominated. Furthermore, if we add a single scalar field then at least at a classical level, minimal coupling is equivalent to conformal coupling under field redefinitions. So this case too may be considered as an example of conformal matter. It is then instructive to extend the discussion to include an arbitrary number of conformally and minimally coupled free scalar fields. 

Since the matter Lagrangian of interest is, by assumption, conformally (Weyl) invariant at a classical level, it makes sense to ``lift" general relativity (GR) to a larger theory possessing the same symmetry. This is done by introducing an extra scalar field which is locally a pure gauge degree of freedom. The full theory is now classically Weyl invariant and it may be viewed with advantage in various Weyl gauges. Its solutions contain all solutions of GR but the theory allows for extended and more general solutions that do not possess a global gauge fixing to GR. In particular, it turns out that while classical cosmological (homogeneous and isotropic) background solutions are typically singular and geodesically incomplete in Einstein gauge (the Weyl gauge in which the gravitational action takes Einstein-Hilbert form, with a fixed Newton's gravitational constant $G$), they are regular and geodesically complete in a more general class of Weyl gauges, where $G$ is no longer constant and can even change sign. In these gauges, generic cosmological background solutions pass smoothly through the ``big bang singularity" and into new regions of field space, including ``antigravity" regions where $G$ is negative~\cite{BST}. Strictly speaking, the lifted classical theory is still incomplete because these highly symmetrical backgrounds are unstable to perturbations in the collapsing phase, leading to diverging anisotropies which cannot be removed via a Weyl gauge choice. It was argued nevertheless that the classical theory possesses a natural continuation across singularities of this kind~\cite{BST2}.

In this paper, following \cite{letter}, we take a different tack.  We ask whether {\it quantum} effects might rescue the theory from its breakdown at big bang-type  singularities.  First, we show that for the simplest types of conformal matter, the Feynman propagator for cosmological backgrounds may be computed exactly, allowing us to explore many issues with precision in this symmetry-reduced (minisuperspace) context. Second, we extend the discussion to include anisotropies, resolving the ordering problems in the quantum (Wheeler-DeWitt) Hamiltonian by imposing covariance under field redefinitions. We go further than the treatment of \cite{letter} by defining the quantum theory along the real axis in superspace, and discovering some remarkable features. For a flat, isotropic universe, quantum effects indeed become large near the singularity. Therefore, strictly speaking, one should not attempt to employ the real classical theory there. Instead, one can solve the quantum Wheeler-DeWitt-type equation for the propagator in complexified superspace, along a contour which avoids the singularity in taking one from the incoming collapsing universe to the outgoing expanding one. Provided the contour stays sufficiently far from the singularity, the semiclassical approximation remains valid all along it, so one can employ {\it complex} solutions of the classical theory to follow the quantum evolution across (or more accurately, around) the singularity. In doing so, we find that while quantum effects are large near the singularity, they take a very special form such that they are ``invisible" in the evolution between incoming and outgoing states. In the final section of this paper, we treat inhomogeneous perturbations, at linear and nonlinear order, showing how they may be followed smoothly and unambiguously across (or, more accurately, around) the big bang singularity, in a similar manner.

The theory we consider consists of Einstein gravity plus radiation and a number of free scalar fields. All of these forms of matter satisfy the strong energy condition, and any cosmological solution necessarily possesses a big bang singularity. However, there is no singularity in the Feynman propagator for closed, open and flat Friedmann-Robertson-Walker (FRW) and for flat, anisotropic Bianchi I cosmologies; the propagator is well behaved across a bounce representing a transition from a large, collapsing classical universe to a  large, expanding one. We also study inhomogeneous perturbations of the isotropic, radiation-dominated universe, studying the propagation of scalar and tensor perturbations across the bounce, at linear and nonlinear order. We find that, in the semiclassical approximation at least and with strictly conformal matter, the incoming vacuum state evolves into the outgoing vacuum state, with no particle production. The bounce may be viewed as an example of quantum-mechanical tunneling, and we use complex classical solutions as saddle points to the path integral, in a manner which generalizes the use of instanton solutions to describe tunneling in more familiar contexts. Some of these results were anticipated in Ref.~\cite{letter}; here we present more details on their derivation, extend them to further cases not discussed in Ref.~\cite{letter}, and provide more mathematical and conceptual background. An alternate interpretation of the antigravity regions, not employing complex solutions, has been presented in Ref.~\cite{Bars}. 

The simplest example of a ``perfect bounce'' is provided by a spatially flat, homogeneous and isotropic FRW universe, filled with a perfect radiation fluid. Adopting conformal time (denoted by $\eta$), {\it i.e.}, choosing the line element to be
\ben
ds^2=a^2(\eta)(-d\eta^2+d\vec{x}^2)\,,
\label{FRW}
\een
one finds the scale factor $a(\eta)\propto\eta$. One way to see this is from the trace of the Einstein equations. For the line element (\ref{FRW}),  $R=6(d^2a/d\eta^2)/a^3$ and for conformal matter, the stress tensor is traceless. So the Einstein equations imply $a(\eta)\propto\eta$. If spatial curvature is included, it is subdominant at small $a$ so that $a$ still vanishes linearly with $\eta$. Hence $a(\eta)\propto\eta$ is a direct consequence of conformal invariance and cosmological symmetry. While the line element (\ref{FRW}) clearly contains a big bang/big crunch singularity at $\eta=0$, it is regular everywhere else, not only along the real $\eta$-axis but also in the entire complex $\eta$ plane. Any complex $\eta$ trajectory that connects large negative and large positive $a$ while avoiding $a=0$ gives a regular, complexified metric that asymptotes to a large, Lorentzian universe in the past and the future, while circumventing the big bang singularity. 

Such complex solutions have been discussed in quantum cosmology for a long time as saddle points of the path integral, for instance in the no-boundary proposal of Hartle and Hawking \cite{nobound,lehners}. The crucial difference in our proposal is that the complex solutions connect two large, Lorentzian regions, identified with a collapsing incoming universe and an expanding outgoing one. The Weyl-invariant lift of GR provides a convenient simplification of the geometry on superspace, cleanly exhibiting its Lorentzian nature and the role of the scale factor $a$ as a single timelike coordinate for both ``gravity" regions. This leads us to a novel formalism for quantum cosmology in which, rather than restricting to positive $a$ and imposing boundary conditions at $a=0$, $a$ is extended to the entire real line. The Feynman propagator turns out to have simple behavior at large negative and positive $a$, describing a contracting and reexpanding universe, respectively, and connecting them through a quantum bounce. The purpose of this paper is to flesh out the details of this formalism, and show how it leads to a novel form of singularity avoidance in the context of an extremely simple (but not altogether unrealistic) cosmology. 

Some of the features of our discussion are not new. The possibility of solving minisuperspace quantum cosmology models exactly by recasting their dynamics as those of a relativistic free particle or harmonic oscillator was pointed out before (see, {\it e.g.}, Refs.~\cite{QCliterature}). However, the crucial new feature in our work is the existence of regular solutions (especially complex ones) that connect two large Lorentzian universes through a quantum bounce. This feature relies on having a positive energy density in radiation, a possibility which, as far as we know, was overlooked in previous work. In fact, our results suggest that the fact that the early Universe was dominated by radiation may be sufficient in itself for a semiclassical quantum resolution of the big bang/big crunch singularity, without the need for less well-motivated ingredients such as exotic forms of matter \cite{matterbounce}, modified theories of gravity \cite{modgravbounce}, or a proposed theory of quantum gravity \cite{qgbounce}. To avoid any potential confusion, let us reemphasize that the theories we consider consist of general relativity with a radiation fluid and a number of free scalars, and nothing more: our use of the Weyl lift of GR does not introduce any additional degrees of freedom.  

The plan of our paper is as follows. In Sec.~\ref{theorysec}, we introduce the Weyl lift of GR plus radiation and scalars, and show how the degree of freedom corresponding to the metric determinant can be isolated straightforwardly, leading us to a Weyl-invariant notion of $a$, the ``scale factor.'' We then study homogeneous, isotropic FRW universes in Sec.~\ref{friedbounce}, showing that in these cases, the Einstein-matter action  corresponds to that of a massive relativistic particle moving in Minkowski spacetime, either freely or subject to a quadratic, Lorentz-invariant potential. We discuss the classical and quantum dynamics of FRW universes, using the classical Hamiltonian analysis to define the Wheeler-DeWitt quantum Hamiltonian. We discuss the Klein-Gordon-type inner product proposed by DeWitt \cite{dewitt}, but take the point of view that the fundamental quantity of interest is really the causal (Feynman) propagator, which is naturally defined as a path integral over four-geometries \cite{teitelboim}. Accordingly, in Sec.~\ref{feynman} we calculate the propagator for various cases of interest. While the Feynman propagator for FRW universes is actually regular at the singularity $a=0$, its asymptotic behavior for large arguments displays interesting pathologies both for closed and, in particular, open universes, so that only flat FRW universes seem to consistently admit a quantum bounce. In Sec.~\ref{anisosec}, we extend the treatment to anisotropic universes of Bianchi I type, including for generality a number of free minimally coupled scalar fields. We resolve the ordering problem in the quantum Hamiltonian, and we are again able to explicitly derive the Feynman propagator. A very special, singular potential arises centered on $a=0$ which, in the minisuperspace context, is harmless and actually invisible in the scattering amplitude between incoming and outgoing states. The coefficient of this singular potential turns out to take a special value for the isotropic universe with zero or one conformally coupled scalars, placing it on the edge of a potential quantum instability, as we discuss. The addition of further conformally coupled scalars moves the theory away from this edge, however.  This is an intriguing result that deserves further attention, as it could be used to select between isotropic and strongly anisotropic universes. In Sec.~\ref{pertsec}, we add inhomogeneities, treated linearly and nonlinearly in the semiclassical approximation where one employs complex classical solutions to the classical Einstein equations. We show how this is sufficient to determine mixing between positive- and negative-frequency modes, and hence to compute the particle production across the ``quantum bounce.'' We find no particle production, but instead verify that the perturbation expansion breaks down at late times due to the formation of shocks in the fluid, a phenomenon which is now physically well understood \cite{fluidpaper}. Section~\ref{conclsec} concludes.

\section{Weyl-invariant cosmology}
\label{theorysec}

We start by studying the cosmology of a universe filled with perfectly conformal radiation and a number $M$ of conformally coupled scalar fields, with gravitational dynamics governed by a lift of GR to a classically Weyl-invariant theory that contains an additional dilaton field $\phi$. We stress again that the field $\phi$ is locally pure gauge, and possesses no nontrivial dynamics of its own. This formalism for GR was developed in Ref.~\cite{BST} and works in any number of dimensions $D>2$ (we will set $D=4$ shortly). The total action we consider is
\bena
S&=&\int d^{D} x \left\{\sqrt{-g}\bigl[\half \left((\partial \phi)^2-(\partial \vec{\chi})^2\right)- \rho\left(\frac{|J|}{\sqrt{-g}}\right)\right.\label{eq1}
\\&&+ \left.\frac{(D-2)}{8(D-1)} (\phi^2 -\vec{\chi}^2)R \bigr]-J^{\mu}\left(\partial_\mu\tilde\varphi+\beta_A\partial_\mu\alpha^A\right)\right\}\,.\nonumber
\eena
The independent dynamical variables are the spacetime metric $g_{\mu\nu}$ (assumed to be Lorentzian throughout), the ``dilaton'' $\phi$, $M$ physical scalar fields $\vec{\chi}=(\chi^1,\dots,\chi^M)$, and a densitized particle number flux $J^\mu$ characterizing the radiation fluid. The latter can be identified by $J^{\mu}=\sqrt{-g}\,n\,U^{\mu}$, where $n$ is the particle number density and $U^\mu$  is the four-velocity vector field satisfying $U^2=-1$; the energy density $\rho$ is only a function of $n$, concretely $\rho(n)\propto n^{\frac{D}{D-1}}$ for radiation which is the case we are interested in. There is a Lagrange multiplier $\tilde\varphi$ which enforces particle number conservation $\partial_\mu J^{\mu}=0$, and $D-1$ further Lagrange multipliers $\beta_A$, with $A=1,\ldots,D-1$, enforcing constraints $J^\mu\partial_\mu\alpha^A=0$ that restrict the fluid flow to be directed along flow lines labeled by the fields $\alpha^A$ which play the role of Lagrangian coordinates for the fluid. 

In general, the fluid energy density $\rho$ would also depend on the entropy per particle. For simplicity, we henceforth assume an isentropic fluid for which this entropy per particle is a constant. The fluid part of our action is then the one given for isentropic fluids in Eq.~(6.10) of Ref.~\cite{Brown}, where further details on the construction of actions of relativistic fluids and their corresponding Hamiltonian dynamics can be found.

The action (\ref{eq1}) is invariant under a Weyl transformation that takes 
\ben
g_{\mu \nu}\rightarrow \Omega^2 g_{\mu \nu}\,,\quad  (\phi, \vec{\chi}) \rightarrow \Omega^{(2-D)/2} (\phi, \vec{\chi})\,,
\label{weyl}
\een
where $\Omega(x)$ is an arbitrary function on spacetime. Such a transformation also takes $\rho\rightarrow\Omega^{-D}\,\rho$. Because of this local conformal symmetry, the field $\phi$ does not correspond to a physical degree of freedom; indeed, if $\phi^2 -\vec{\chi}^2>0$ everywhere, one can gauge fix the conformal symmetry to recover the usual Einstein-Hilbert formulation of GR. It is then clear that there is no physical ghost in the theory even though $\phi$ appears in Eq.~(\ref{eq1}) with the wrong-sign kinetic term. 

Let us make this explicit. For $\phi^2 -\vec{\chi}^2>0$, one can go to ``Einstein gauge'' by performing a conformal transformation (\ref{weyl}) that takes
\ben
\phi^2 -\vec{\chi}^2 \rightarrow \const =: \frac{D-1}{2(D-2)\pi G}
\label{transf}
\een
where $G$ is Newton's constant. Note that Eq.~(\ref{transf}) does not entirely fix the gauge freedom as one can still perform a global rescaling that takes the constant to a different one; the exact value of Newton's constant is arbitrary and corresponds to a choice of units. Einstein gauge corresponds to constraining the $(M+1)$-vector formed by $(\phi,\vec{\chi})$ to a hyperboloid $H^M$ in $(M+1)$-dimensional field space at each point in spacetime. One can introduce an explicit parametrization of this hyperboloid by $M$ coordinates $\nu^i$,
\ben
\phi=\phi(\nu^1,\ldots,\nu^M)\,,\quad\chi^i=\chi^i(\nu^1,\ldots,\nu^M)\,,
\een
so that in this gauge the action (\ref{eq1}) reads
\bena
S&=&\int d^{D} x \left\{\sqrt{-g}\bigl[-\half G_{ij}(\nu)\,\partial \nu^i\cdot\partial\nu^j- \rho\left(\frac{|J|}{\sqrt{-g}}\right)\right.\nonumber
\\&&+ \left.\frac{1}{16\pi G} R \bigr]-J^{\mu}\left(\partial_\mu\tilde\varphi+\beta_A\partial_\mu\alpha^A\right)\right\}\,,\label{newaction}
\eena
where $G_{ij}(\nu)$ is a positive definite metric of constant negative curvature on the gauge-fixed field space parametrized by the $\nu^i$. Again, Eq.~(\ref{newaction}) shows that there are no physical ghosts in the theory, at least as long as $\phi^2-\vec\chi^2>0$. 

There are two different sectors in the space of field configurations where Einstein gauge is available, corresponding to ``future-directed'' and ``past-directed'' (in field space) configurations, {\it i.e.}, to $\phi>0$ or $\phi<0$. There are also regions where $\phi^2-\vec\chi^2$ becomes negative, identified with ``antigravity'' in Ref.~\cite{BST} as they would appear to correspond to a negative $G$. We identify such regions with imaginary values of the scale factor and show how the passage of the Universe through antigravity regions is a semiclassical representation of what is really a quantum bounce, similar to how quantum tunneling can be described by complex classical trajectories. The antigravity regions do contain a ghost, as now $(\phi,\vec{\chi})$ would be constrained to de Sitter space ${\rm dS}^{M-1,1}$ which has a timelike direction. These regions and their ghost excitations do not appear in the physical ``in" and ``out" states of the theory, which are defined in asymptotic timelike regions where $\phi^2-\vec\chi^2\rightarrow\infty$; nevertheless, the existence of these regions can cause pathologies in the quantum theory if initial gravity states can propagate into the antigravity regions, as we will see in Sec.~\ref{feynfrw}.

Setting $D=4$, to make this more precise it is now useful to define a scale factor, or rather its square $a^2$, with the following properties: it should be Weyl invariant, so that it takes the same value in any conformal gauge. It should respect the $O(M,1)$ isometry of the metric on the space of scalar fields (defined by the kinetic terms) and so depend only on the combination $\phi^2-\vec{\chi}^2$, the radiation density $\rho$ and the metric determinant $g$. It should have physical dimensions of an area (in the usual conventions $\hbar=c=1$), and scale like the square of the scale factor for an FRW universe in Einstein gauge in conformal time. These properties fix the ``squared scale factor,'' up to an overall constant, to be
\ben
a^2\equiv \frac{1}{2\rho}(-g)^{-\frac{1}{4}}\left(\phi^2-\vec{\chi}^2\right)\,.
\label{scalefactor}
\een
We use Eq.~(\ref{scalefactor}) as a natural definition in general gauges (the motivation for the factor $\half$ becomes clear shortly). Note that $a^2$ is in general not positive; if we assume positive $\rho$ and a Lorentzian metric, as we always do in the following, then in the antigravity regions $a^2<0$ and so $a$ is imaginary. This definition of $a$ differs from the one in Ref.~\cite{BST} as it depends on the energy density of the radiation. In Ref.~\cite{BST}, there was no such dependence. Instead, factors of Newton's constant were used to ensure the correct physical dimensions.

\begin{figure}[htp]
\includegraphics[scale=1]{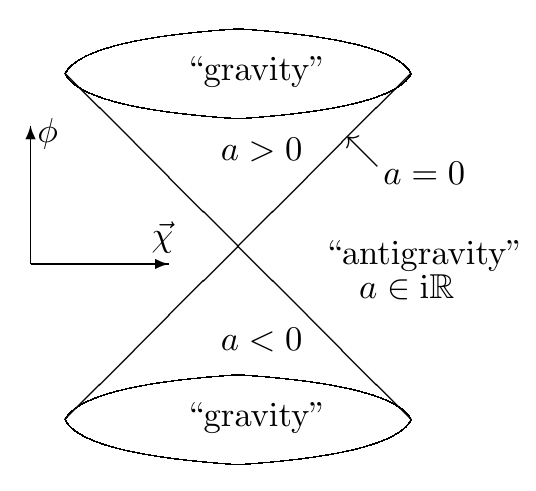}
\caption{Associating positive, negative and imaginary scale factor $a$ to different regions in field space. The light cone corresponds to the singularity $a=0$.}
\label{scalefactorim}
\end{figure}

For $a^2>0$, we fix the sign of $a$ by choosing a time orientation in field space: $a$ is defined to have the same sign as $\phi$. The Minkowskian field space parametrized by $(\phi,\vec\chi)$ is then partitioned into two regions with real $a$ and one region with imaginary $a$; see Fig.~\ref{scalefactorim}. Note that the entire light cone corresponds to $a=0$. This picture, as we have anticipated, gives physical meaning to positive, negative and imaginary $a$, generalizing the case of pure radiation, $M=0$, where there are no spacelike directions and $a$ takes values along the real axis, as in the example discussed in the introduction.

Apart from Einstein gauge, another gauge that we will often employ, again following the framework introduced in Ref.~\cite{BST}, is ``Weyl gauge'' where the metric determinant $g$ is fixed to a constant (typically $-1$). This gauge is available whenever the metric is nonsingular; in particular, it covers the entire field space pictured in Fig.~\ref{scalefactorim}, encompassing both gravity and antigravity regions. It is often a convenient gauge to work in. In Weyl gauge, the expression for the scale factor reduces to $a^2=\frac{1}{2\rho}\left(\phi^2-\vec{\chi}^2\right)$, where for homogeneous models by energy-momentum conservation $\rho$ is constant. $a$ is then proportional to the (signed) timelike $O(M,1)$-invariant distance from the origin in field space, making it a natural choice of time coordinate on superspace.

For highly symmetric solutions such as FRW universes, conformal symmetry can be used to eliminate curvature singularities in the metric by moving them into 0's of the quantity $\left(\phi^2-\vec{\chi}^2\right)$. Since this quantity has no geometric interpretation, it is {\it a priori} reasonable for it to vanish or change sign. However, following the dynamical evolution through such points is in general problematic because the effective Newton constant diverges so gravity becomes strongly coupled. This is reflected, for example, in the behavior of tensor (gravitational wave) perturbations, which diverge as the effective Newton constant does. This leads to a diverging Weyl curvature which cannot be removed because it is conformally invariant. Nevertheless, in the presence of scalar fields (such as the electroweak Higgs boson) there is generically no Mixmaster chaos and one expects the classical evolution to become ultralocal and Kasner-like. There are a number of asymptotically  conserved classical quantities, including the Kasner exponents, suggesting a natural matching rule across the singularity~\cite{BST2} but the issue has not been conclusively settled~\cite{debate}.  

In this paper, we take a different approach. We show that by extending the classical discussion to a quantum picture one can avoid the critical surface $a=0$ where the theory becomes problematic, independently of any Weyl gauge choice. We give a description of nonsingular quantum bounces in terms of analytic continuation in $a$, where the Universe evolves from large negative $a$ to large positive $a$ along a contour in the complex $a$-plane which avoids $a=0$. We argue that as long as the quantum mechanics of the $a$ degree of freedom make sense, the classical singularity at $a=0$ can be avoided without obstruction.

\section{FRW bounces}
\label{friedbounce}

As a first step, we perform the familiar symmetry reduction of our theory to homogeneous and isotropic FRW universes, with the metric assumed to be of the form
\ben
ds^2=A^2(t)(-N^2(t)dt^2+h_{ij}dx^idx^j)
\een
where $h_{ij}$ is a fixed metric on hypersurfaces of constant $t$, which has constant three-curvature $R^{(3)}=6\kappa$. Note our use of a conformal lapse function $N$; the usual definition of the lapse would be $N_0(t)=A(t)\cdot N(t)$. We can now set the function $A(t)$ to one by a conformal transformation, so that the metric becomes nondynamical and all dynamics are in the scalar fields $\phi$ and $\vec\chi$. Also, with FRW symmetry $J^\mu=\sqrt{h}n\,\delta_0^\mu$, and the action (\ref{eq1}) reduces to
\ben
S=V_0\int dt\,\bigl[\frac{\dot{\vec{\chi}}^2-\dot{\phi}^2}{2 N}+N\left(\frac{\kappa}{2}(\phi^2-\vec{\chi}^2) -\rho(n)\right)-\tilde\varphi \dot{n}  \bigr]\,,
\label{eq2}
\een
where $\dot{}$ denotes derivative with respect to $t$ and $V_0=\int d^3 x\sqrt{h}$ is the comoving spatial volume (which, as usual for minisuperspace models, must be assumed to be finite). We have simplified the last term including the Lagrange multipliers which would be $-n(\dot{\tilde\varphi}+\beta_A\dot\alpha^A)$ since the equations of motion involving $\beta_A$ and $\alpha^A$ are clearly redundant in FRW symmetry. As before, $\rho(n)\propto n^{4/3}$, and we can replace $n$ by $\rho$ as the independent variable.

\begin{figure}[htp]
\includegraphics[scale=1]{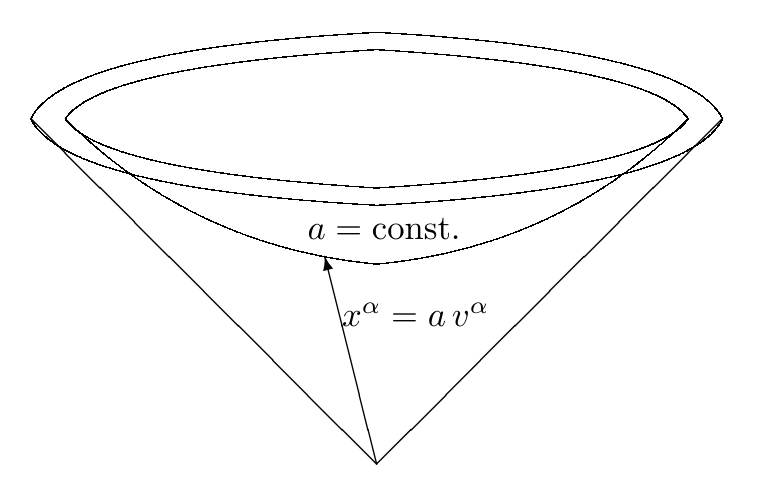}
\caption{The points of constant (real) $a$ form a hyperboloid parametrized by $v^\alpha$.}
\label{hyperbfig}
\end{figure}

It is evident that Eq.~(\ref{eq2}) is the action for a relativistic massive free particle (for $\kappa=0$) or a relativistic massive particle in a harmonic potential or a harmonic ``upside-down'' potential (for $\kappa\neq 0$) moving in $(M+1)$-dimensional Minkowski spacetime. To make this more explicit, we can introduce new variables
\ben
x^\alpha:=\frac{1}{\sqrt{2\rho}}(\phi,\vec\chi)\,,\;\alpha=0,\ldots,M\,,\quad m:=2V_0\rho\,,
\label{newvariables}
\een
so that Eq.~(\ref{eq2}) now takes the form
\ben
S= \int dt \bigl[\frac{m}{2}\left(\frac{1}{N}\dot{x}^\alpha \dot{x}_\alpha-N(\kappa \,x^\alpha x_\alpha+1)\right)-\varphi\dot{m}\bigr]
\label{eq3}
\een
where we have redefined the Lagrange multiplier for simplicity, $\varphi:=\tilde\varphi V_0 (dn/dm)$, and the Minkowski metric on the space of scalar fields $\eta_{\alpha\beta}={\rm diag}(-1,1,1,\ldots)$ is used to raise and lower indices. A crucial role is played by the mass $m$ which corresponds to (twice) the total energy in the radiation; the limit $m\rightarrow 0$ would correspond to a massless relativistic particle moving in a potential, which is the case well known in minisuperspace quantum cosmology with scalar fields \cite{QCliterature}. Having a positive mass, and hence timelike trajectories as classical solutions, is one of the essential features of our model that leads to a bounce. With the definition (\ref{newvariables}), the Weyl-invariant scale factor is simply $a^2=-x^2$, which explains the factor $\half$ in Eq.~(\ref{scalefactor}). The variable $a$ is simply a time coordinate on superspace. One can introduce it explicitly by setting
\ben
x^\alpha=a\,v^\alpha\,,\quad v^2=-1
\label{vparam}
\een
so that $v^\alpha$ is restricted to a hyperboloid $H^M$ (see Fig.~\ref{hyperbfig}).
This parametrization, which isolates the physical scalar fields as the variables $v^\alpha$, is useful below.

Starting from Eq.~(\ref{eq3}), the classical equations of motion are
\bena
&&\frac{1}{N} \frac{d}{dt}\left(\frac{m\dot{x}^{\alpha}}{N}\right)+m\kappa x^\alpha =0\,,
\label{eom1}
\\&&\frac{1}{N^2}\dot{x}^{\alpha}\dot{x}_{\alpha}+\kappa x^{\alpha}x_{\alpha}=-1\,,
\\&&\dot m =0\,,\quad \dot\varphi= -\frac{1}{2N}\dot{x}^{\alpha}\dot{x}_{\alpha}+\frac{N}{2}\left(\kappa x^{\alpha}x_{\alpha}+1\right)\label{lasteom}\,.
\eena
The general solution to these equations is $m=\const$,
\bena
x^{\alpha}(t)&=&\frac{x_1^\alpha}{\sqrt{\kappa}} \exp\left(\im\,\sqrt{\kappa}\int\limits_0^t dt'\,N(t')\right)\nonumber
\\&&+\frac{x_2^\alpha}{\sqrt{\kappa}} \exp\left(-\im\,\sqrt{\kappa}\int\limits_0^t dt'\,N(t')\right)
\label{classsol}
\eena
with $x_1\cdot x_2=-\frac{1}{4}$; the lapse function $N(t)$ is arbitrary and $\varphi(t)$ is determined from integrating Eq.~(\ref{lasteom}). For $\kappa=0$, the general solution is simply a general timelike straight line in Minkowski spacetime,
\ben
x^\alpha(t)=x_1^\alpha \int\limits_0^t dt'\,N(t') +x_2^\alpha
\een
with $x_1^2=-1$. For $\kappa=0$, all solutions describe a bounce, similar to the example in the introduction: the Universe comes in from negative real infinite $a$, goes through $a=0$ followed in general by an ``excursion'' into imaginary $a$, and crosses $a=0$ again before going off to real positive infinity. When we go quantum, since the action is quadratic, the saddle point approximation is exact and the quantum dynamics is given purely in terms of these classical solutions. When viewed as saddle points, these trajectories can be deformed in the complex $a$-plane so that the singularity $a=0$ is avoided. The situation is more subtle for $\kappa<0$, where there are spacelike as well as timelike solutions, and for $\kappa>0$ where there is a turnaround in the classical solutions and the Universe must recollapse due to the spatial curvature. In Sec.~\ref{feynman}, we see how the more complicated structure of solutions for $\kappa\neq 0$ is reflected in a pathological behavior of the Feynman propagator for large arguments.

\subsection{Canonical formalism}
\label{canonical}

In order to pass to the Hamiltonian formalism, following Dirac's algorithm \cite{dirac}, one computes the canonical momenta for the action (\ref{eq3}) and finds
\ben
p_\alpha=\frac{\partial \mathcal{L}}{\partial \dot{x}^\alpha}=\frac{m}{N}\dot{x}_{\alpha}\,,\quad p_m\approx -\varphi\,, \quad p_N\approx 0\,, \quad p_\varphi \approx 0\,. 
\een
While the first equation can be inverted to express the velocities $\dot{x}^{\alpha}$ in terms of the momenta $p_\alpha$, the last three equations are {\em primary constraints}---we use Dirac's notion of ``weak equality'' $\approx$ for equations that hold on the constraint surface. The second and fourth constraint would be second class, meaning one has to introduce a Dirac bracket and ``solve'' them. However, in this case, one can use the shortcut of simply identifying $-\varphi$ with the momentum conjugate to $m$ and removing the separate variable $p_\varphi$. This is equivalent to saying that the term $-\varphi\dot{m}$ in Eq.~(\ref{eq3}) is part of the symplectic form $p_i\dot{q}^i$ so that one can read off $p_m=-\varphi$.
\\The Hamiltonian is then
\ben
\mathcal{H}=N\left(\frac{p^2}{2m}+\frac{m}{2}(\kappa x^2+1)\right)+\xi\, p_N\,.
\label{frwhamilt}
\een
Preservation of $p_{N}\approx 0$ under time evolution gives the secondary, Hamiltonian constraint, 
\ben
C:=\frac{p^2}{2m}+\frac{m}{2}(\kappa x^2+1)\approx 0\,.
\een 
$N$ can then be treated as a Lagrange multiplier; it only enters linearly in the Hamiltonian, and its time evolution under $\mathcal{H}$ is $ \dot{N}=\{N,\mathcal{H}\}=\xi$ where $\xi$ is  undetermined. Removing $(N,p_{N})$ from the phase space (and setting $\xi=0$ in the Hamiltonian), we are left with the canonical pairs $(x^\alpha, p_\alpha)$ and $(m,p_m)$, subject to the constraint $C$, which trivially satisfies $\{C,\mathcal{H}\}=0$. $C$ generates time reparametrizations,
\bena
&&\delta_N x^{\alpha}=\{x^\alpha,NC\}=\frac{Np^{\alpha}}{m}\,,
\\&&\delta_N p_{\alpha}=-Nm\kappa x_{\alpha}\,,
\\&&\delta_N p_m=-N\left(-\frac{p^2}{2m^2}+\half(\kappa x^2+1)\right)\,,
\eena
which correspond to the Lagrangian notion of time reparametrization, by the equations of motion (\ref{eom1})--(\ref{lasteom}).

\subsection{Quantization}

Having set up the canonical formalism, we can proceed with quantization in the standard way. The Hamiltonian constraint is imposed as an operator equation  restricting the set of physical states. In the $(x,m)$ representation for the wave function, this is the Wheeler-DeWitt equation
\ben
\frac{1}{2m}\left(-\Box_x+m^2(\kappa x^2+1)\right)\Psi(x,m)=0.
\label{wdw1}
\een
Different $m$ sectors simply decouple, as a consequence of conservation of the total energy in the radiation, with no transitions between different $m$ values allowed. An alternative representation of wave functions is obtained by separating the scale factor from the physical scalar field degrees of freedom, as in Eq.~(\ref{vparam}), and introducing a set of coordinates $\nu^i$, $i=1,\ldots,M$, on the hyperboloid $H^M$. The Wheeler-DeWitt equation then becomes
\ben
\left(\frac{\partial^2}{\partial a^2}+\frac{M}{a}\frac{\partial}{\partial a}-\frac{1}{a^2}\Delta_{H^M}+m^2(1-\kappa a^2)\right)\Psi(a,\nu^i,m)=0
\label{wdw2}
\een
where $\Delta_{H^M}$ is the Laplace-Beltrami operator on $M$-dimensional hyperbolic space, {\it i.e.}, on the space parametrized by the coordinates $\nu^i$. For example, using Beltrami coordinates $\nu^i=x^i/x^0$ one would have
\ben
\Delta_{H^M}=(1-\vec\nu^2)\left[\left(\delta^{ij}-\nu^i\nu^j\right)\partial_i\partial_j-2\nu^i\partial_i\right]\,.
\een
This coordinate choice on superspace makes the role of the timelike coordinate $a$ explicit. The Wheeler-DeWitt equation can then be simplified by Fourier transforming from the $\nu^i$ coordinates to their conserved momenta $\zeta$,
\ben
\left(\frac{\partial^2}{\partial a^2}+\frac{M}{a}\frac{\partial}{\partial a}-\frac{c}{a^2}+m^2(1-\kappa a^2)\right)\Psi(a,\zeta^i,m)=0
\label{Wdwfrw}
\een
with 
\ben
c\equiv -\frac{1}{4}(M-1)^2-\vec\zeta^2
\label{cpminimal}
\een
corresponding to the eigenvalues of the Laplacian on $H^M$ (for $M\ge 1$). As we see in Sec.~\ref{anisosec} below, the same form of the Wheeler-DeWitt equation applies when including anisotropies in a Bianchi I model or additional minimally coupled scalar fields, with the only change that the constant $c$ receives additional contributions from conserved anisotropy and scalar field momenta as well as from fixing ordering ambiguities.

A natural inner product on solutions of a second-order equation like Eq.~(\ref{wdw2}) is the Klein-Gordon-like norm
\bena
\langle\Psi|\Phi\rangle&\equiv&\im a^M\int d^M\nu\,dm\;\sqrt{g_{H^M}}\left(\Psi^*(a,\nu^i,m)\frac{\partial}{\partial a}\Phi(a,\nu^i,m)\right.\nonumber
\\&&-\left.\frac{\partial}{\partial a}\Psi^*(a,\nu^i,m)\Phi(a,\nu^i,m)\right)\,,
\label{norm}
\eena
with $g_{H^M}$ being a constant negative curvature metric on $H^M$, which is conserved under time evolution, {\em i.e.} independent of $a$, for solutions of Eq.~(\ref{wdw2}). This inner product was introduced by DeWitt \cite{dewitt} and has the well-known problem (if it is used to define probabilities) that it is only positive on positive-frequency solutions to Eq.~(\ref{wdw2}), when they exist. For some simple cosmological models, this subspace is well defined, and may be interpreted as the space of expanding quantum universes: if $a$ is taken to be positive, a wave function describing an expanding universe must be an eigenstate of $p_a=-\im \partial_a$ with negative eigenvalue (note that $p_a=-m\dot{a}/N$ and so $\dot{a}>0$ means $p_a<0$), {\it i.e.}, a positive-frequency solution. This  notion of positive frequency breaks down for cosmological models with recollapsing solutions, such as the FRW universe with $\kappa>0$, where it is only well defined until one reaches the turning point, and it is known that a decomposition into positive and negative frequencies of the type we are using here is not available in general \cite{kuchar}. The question of how to define meaningful probabilities in quantum cosmology has, of course, been a matter of long debate (see, {\it e.g.}, Refs.~\cite{nobound,halliwell,wiltshire}). 

We do not aim to resolve this debate here. The only use we make of the DeWitt norm (\ref{norm}) is to help us construct the Feynman propagator from mode function solutions of the Wheeler-DeWitt equation. The expansion rate $-p_a$ does play the role of an energy, which leads us to adopt Feynman's picture for quantum field theory in which positive energy ({\it i.e.}, expanding) states are propagated forward in proper time. The natural two-point function we consider below in Sec.~\ref{feynman} is hence the Feynman propagator. In what follows we alternate between the path integral and the  Feynman propagator as basic formulations of quantum cosmology, explicitly showing their equivalence in simple cases.

So far, this is a completely standard definition of a minisuperspace model in Wheeler-DeWitt quantum cosmology. However, there is one crucial difference between our approach and previous treatments, in that we do not restrict the wave function to positive $a$, nor do we impose any boundary condition at $a=0$ [such as the popular choice $\Psi(a=0)=0$]. At fixed $m$, the domain of the wave function is simply $\mathbb{R}^{M,1}$. Any boundary condition at $a=0$ would seem artificial from the viewpoint of classical solutions such as classical FRW ``bounces" which connect negative and positive $a$, as we have described, and is also generally inconsistent with the wave function describing an expanding Universe, {\it i.e.}, a positive-frequency solution. In our proposal, the natural choice of wave functions corresponds to positive-frequency solutions that asymptote to plane waves at large $|a|$, where the Universe becomes semiclassical, while we allow for irregular behavior in the wave function at $a=0$. The examples we consider all admit a semiclassical Wentzel-Kramers-Brillouin (WKB) expansion in which one can deform the contour from the real $a$-axis into complex $a$, avoiding $a=0$ entirely.

\section{Feynman Propagator for FRW Universes}
\label{feynman}

The Feynman propagator is one of the most basic ingredients in relativistic quantum theory. In quantum gravity, it plays the role of a causal Green's function for the Wheeler-DeWitt equation, arising from a path integral in which one integrates only over positive values of the lapse function \cite{teitelboim}. If one considers amplitudes in which $a$ changes sign, as we do, then the Feynman propagator takes one from a contracting universe in the initial state to an expanding one in the final state, via a singularity of the big bang type. Such an amplitude provides a natural way to describe the ``emergence" of spacetime from quantum-mechanical first principles~\cite{feldbrugge}.

In this section, we show how to calculate the Feynman propagator for FRW universes directly from the path integral; in particular, the path integral may be used to define the propagator without the need for an additional ``$\im \epsilon$" prescription and, furthermore, the propagator directly defines the positive- and negative-frequency vacuum modes. As we have stressed, with FRW symmetry the action is quadratic and the saddle-point approximation is therefore exact for the path integration over the phase-space variables. However, we also have a constraint (the Friedmann equation) which must be imposed via an additional integration over the lapse function or Lagrange multiplier $N$ \cite{hennteit}. This integral is no longer Gaussian and has to be considered with more care. Things are considerably simpler for the flat FRW case $\kappa=0$, where we have seen that the dynamics are just those of a massive relativistic free particle in Minkowski spacetime. We therefore begin by reviewing how the Feynman propagator for a relativistic particle is calculated both from a path integral and as a Green's function for the differential equation satisfied by physical wave functions. We then extend these methods to treat general FRW universes for the types of matter we consider.

\subsection{Relativistic particle}
\label{relpart}

Consider the action for a relativistic massive particle in $(M+1)$-dimensional Minkowski spacetime, 
\ben
{\cal S}=\frac{m}{2} \int dt \left(\frac{\dot{x}^\alpha \dot{x}^\beta\eta_{\alpha \beta}}{N}-N\right)\,,
\label{e1}
\een
where $m>0$, $x^\alpha(t)$ is the parametrized particle world line and $N(t)$ is the ``einbein.'' Classically,  $N$ may be eliminated using its equation of motion, obtaining the manifestly reparametrization-invariant action $S=-m\int dt \sqrt{-\dot{x}^2}$. Quantum mechanically, it is more convenient to fix the reparametrization invariance [see also the discussion of gauge fixing below Eq.~(\ref{phsppathint})]: one can work in a gauge in which $t$ varies over a fixed range, conveniently taken to be $-\half <t<\half$, and $N$ is a $t$-independent constant, equal to the total, reparametrization-invariant time $\int dt \,N$ which we call $\tau$. The Feynman propagator is then given by the path integral
\bena
G(x| x') &=&\int  d\tau\,\mathcal{D}x\,  \exp\left[ \im \frac{m}{2} \int_{-\half}^\half dt \left(\frac{\dot{x}^2}{\tau} -\tau \right)\right] \nonumber
\\&=&\im\int_0^{\infty}  d\tau \left(\frac{m}{2\pi\im\tau}\right)^{\frac{M+1}{2}}e^{-\im \frac{m}{2}\left(\frac{\sigma}{\tau}+\tau\right) }\,,\label{e2}
\eena
where $\sigma\equiv -(x-x')^2$ and $\tau$ should be integrated over positive values. Evaluating the Gaussian path integrals is straightforward, with the only unusual factor being the prefactor of $\im$ in the second line, which arises from the functional integral over $x^0$, which has the ``wrong sign'' kinetic term so that the overall phase factor contributed is $e^{+\im \pi/4}$ rather than the usual $e^{-\im \pi/4}$. 

The second line of Eq.~(\ref{e2}) is, of course, just the familiar Schwinger representation of the Feynman propagator, in which the exponent is the classical action evaluated on a solution of the equations of motion $\ddot{x}=0$, satisfying the correct boundary conditions, {\it i.e.},  $x(t)=x(t+\half)+x'(\half-t)$, and the prefactor is given by the usual (regularized) functional determinant. Note that this solution is only the saddle point for the functional integral over paths $x(t)$, at fixed $\tau$. The constraint $\dot{x}^2=-\tau^2$ arises subsequently, as the condition for a saddle point in the exponent of the $\tau$ integral. In fact, once the saddle point is chosen, the integration contour is then fixed (up to an equivalence class of contours yielding the same result) as the complete extension of the corresponding steepest descent contour. Integrating over negative proper time in Eq.~(\ref{e2}) would reverse the notion of time ordering, whereas integrating over both positive and negative $\tau$ would lead to a symmetrized two-point function in which one sums over both time orderings, {\it i.e.}, the Hadamard propagator.

For $M>0$, the $\tau$ integral in Eq.~(\ref{e2}) has a potential divergence at $\tau=0$. In fact, the integral converges at all real values of $\sigma$ except $\sigma=0$. That divergence is real: the Feynman propagator is singular for null-separated points. For other real values of $\sigma$, given that the integral converges for all $\sigma$ in the lower-half $\sigma$ plane, one may define the Feynman propagator as the boundary value of the function defined by the integral, which is analytic in the lower-half $\sigma$ plane. Traditionally, the mass $m$ is also taken to have a small negative imaginary part, in order to make the $\tau$ integral absolutely convergent at large positive values (Feynman's ``$\im\epsilon$" prescription). Equivalently, one may distort the $\tau$-contour to run to infinite values below the real axis. The integral (\ref{e2}) may be directly expressed in terms of a Hankel function, whose properties confirm these general arguments (see the appendix). 

\begin{figure}[htp]
\includegraphics[scale=0.25]{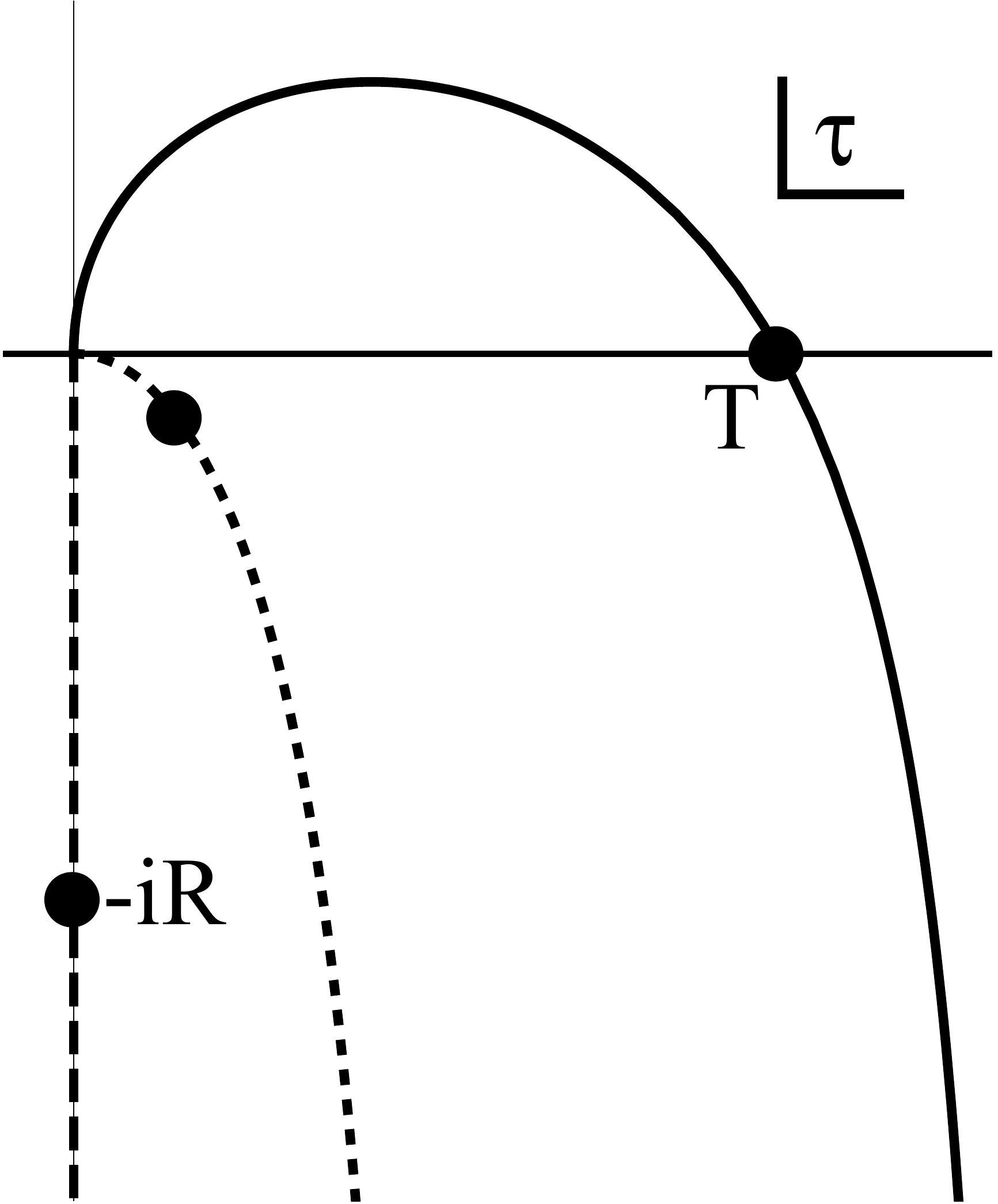}
\caption{Integration contours in the complex $\tau$ plane, for the relativistic massive propagator, defined in Eq.~(\ref{e2}). As $\sigma \equiv -(x-x')^2$ is varied, from timelike values $\sigma=T^2$ with $T$ real, to spacelike values $\sigma=-R^2$, with $R$ real, by passing beneath the origin in the complex $\sigma$ plane, the saddle point in $\tau$, shown by the black point, moves correspondingly. In each case, the integral is defined by the associated steepest descent contour running from $0$ to $\infty$, also shown.} 
\label{conts}
\end{figure}

It is instructive, however, to evaluate the $\tau$ integral in Eq.~(\ref{e2}) in the saddle-point approximation. First, consider timelike separations,  $\sigma =T^2>0$. The exponent in the $\tau$ integral (\ref{e2}) is stationary at $\tau=+T$ and $\tau=-T$, but only the former saddle point is relevant to the contour we want, which is deformable into the positive $\tau$-axis. The saddle point at $\tau=+T$  gives rise to a positive-frequency result, $G \sim e^{-\im m T} $ at large $T$. The integration contour for $\tau$ may then be taken to be the corresponding steepest descent contour, shown as the solid line in Fig.~\ref{conts}. Now consider analytically continuing $T$ through the lower right quadrant to the negative imaginary axis, $T\rightarrow-\im R$. It follows that $G$ converges as $G\sim e^{-m R}$ at large $R$. Correspondingly, this continuation implies that $\sigma=T^2$ runs {\it below} the origin in the complex $\sigma$ plane to negative values.  The corresponding saddle point for the $\tau$ integral moves as shown in Fig.~\ref{conts}, passing below the origin in the complex $\tau$ plane and down the imaginary $\tau$-axis. At  spacelike separations, $\sigma=-R^2<0$, the saddle point is at $\tau=-\im  R$, and the propagator falls exponentially with spacelike separation. Notice that although the classical constraint $\dot{x}^2=-\tau^2$ remains satisfied at the saddle point, the saddle-point value for $\tau$ is imaginary, and hence classically disallowed. Hence, the propagation of a massive relativistic particle in spacelike directions may be viewed as a semiclassical quantum tunneling process, mediated by a complex classical solution. 

In order to match the path-integral definition of the Feynman propagator to its definition as a Green's function, it is convenient to use a time slicing with
maximal spatial symmetry. Here, the trivial time slicing defined by $x^0$ is spatially homogeneous, so one can Fourier transform in the spatial coordinates and reduce the problem to a single timelike dimension. Defining
\ben
G(x|x')=\int \frac{d^M\vec{k}}{(2 \pi)^M}\, e^{\im \vec{k}\cdot (\vec{x}-\vec{x}')} G_k(x^0,x'^0)\,,
\label{e6}
\een
one finds that $G_k(x^0,x'^0)$ is given by
\ben
 G_k(x^0,x'^0) = \im\int_0^{\infty}  d\tau \sqrt{\frac{m}{2\pi\im\tau}}\,e^{-\im \frac{m}{2}\left(\frac{(x^0-x'^0)^2}{\tau}+\frac{\omega_k^2}{m^2}\tau\right) }
\label{fourierint}
\een
where $\omega_k\equiv \sqrt{k^2+m^2}$.

By considering the saddle-point approximation to Eq.~(\ref{fourierint}), we see that the Fourier-transformed Feynman propagator asymptotically satisfies
\bena
G_k(x^0,x'^0)&\sim& e^{-\im  \omega_k x^0}\,,\quad x^0\rightarrow +\infty\,,\; x'^0\;{\rm fixed},
\label{asymp1}
\\G_k(x^0,x'^0)&\sim& e^{+\im  \omega_k x'^0}\,,\quad x'^0 \rightarrow -\infty\,,\; x^0\;{\rm fixed}.
\label{asymp2}
\eena
Such asymptotic expressions can be used to fix boundary conditions for the corresponding wave (Wheeler-DeWitt) equation, as we do shortly.

Formally, $G(x| x')$ is the matrix element $\langle x|\int_0^{\infty} d\tau\; e^{-\im H \tau}|x'\rangle = -\im \langle x| H^{-1}|x' \rangle$, where we again assume the integral converges at infinite $\tau$. Hence, suitably defined,  $G(x|x')$ should obey 
\ben
H_x G(x| x')=-\im \delta^{M+1}(x-x'), \label{e3}
\een
where $H_x$ is the Hamiltonian in the $x$-representation, $H_x= \frac{1}{2 m} (-\Box_x+m^2)$.
We can check Eq.~(\ref{e3}) is indeed satisfied by applying $H_x$ to the last line of Eq.~(\ref{e2}) and using the fact that the integrand is a product of free-particle quantum-mechanical propagators,
\bena
&&\frac{1}{2 m} (-\Box_x+m^2)G(x|x')\nonumber
\\&=&\im\int_0^{\infty} d\tau \,\im \frac{d}{d\tau} \left\{  \left(\frac{m}{2\pi\im\tau}\right)^{\frac{M+1}{2}}e^{\frac{\im m}{2}\left(\frac{(x-x')^2}{\tau}-\tau\right) }\ \right\}\nonumber
\\& =&\lim_{\tau \to 0} \left(\frac{m}{2\pi\im\tau}\right)^{\frac{M+1}{2}}e^{\frac{\im m}{2}\left(\frac{(x-x')^2}{\tau}\right)},
\label{e5}
\eena
where the limit should be taken along the appropriate contour in the complex $\tau$ plane. The last line of Eq.~(\ref{e5}) is a representation of the $(M+1)$-dimensional delta function: separating it into a product of similar terms for each coordinate $x^\alpha$, we  determine the coefficient of the corresponding delta function by Fourier transforming with respect to $x^\alpha$, and then taking the limit $\tau \rightarrow 0$. For the timelike  coordinate we  obtain $-\im\delta(x^0-x'^{0})$, whereas for the spacelike coordinates we obtain $\delta^M(\vec{x}-\vec{x}')$. Together, these results verify Eq.~(\ref{e3}).

The point is now that the Feynman propagator can also be computed by directly solving Eq.~(\ref{e3}) in terms of mode functions, again because the Fourier transform allows reduction of the problem to one dimension. Writing both the delta function and the propagator in Eq.~(\ref{e3}) as Fourier transforms, one sees that $G(x|x')$ clearly satisfies Eq.~(\ref{e3}) as long as  
\ben
(\partial_0^2 +\omega_k^2)G_k(x^0,x'^0)=-2\im m \delta(x^0-x'^0)\,.
\een This equation is solved by
\bena
G_k(x^0,x'^0)&=&-\frac{2\im m}{W(\psi^k_1,\psi^k_2)} \left(\psi^k_1(x'^0)\psi^k_2(x^{0}) \theta(x^0-x'^0)\right.\nonumber
\\&&\left. +\psi^k_1(x^0)\psi^k_2(x'^{0}) \theta(x'^0-x^0)\right)\,,
\label{e7}
\eena
where $\psi^k_1$ and $\psi^k_2$ are two independent solutions to the homogeneous equation $(\partial_0^2 +\omega_k^2)\psi=0$, and $W(\psi_1,\psi_2)= \psi_1\psi_2' -\psi_2 \psi_1'$ is the natural conserved ({\it i.e.}, $x^0$-independent) inner product, or Wronskian. The dependence of the Feynman propagator at large positive and negative times now determines the appropriate choices for $\psi^k_1(x^0)$ and $\psi^k_2(x^0)$: comparing Eqs.~(\ref{asymp1}) and (\ref{asymp2}) with Eq.~(\ref{e7}) we infer that, up to irrelevant constants,  $\psi^k_1= e^{+\im \omega_k x^0}$ and $\psi^k_2= e^{-\im \omega_k x^0}$. Inserted into Eqs.~(\ref{e6}) and (\ref{e7}),  these give the usual expression for the Feynman propagator in ``time-ordered'' form. This shows how, in the example of the relativistic particle, the correct boundary conditions that define the Feynman propagator as one particular solution of Eq.~(\ref{e3}) can be obtained from the asymptotic behavior of its path-integral definition. We will now proceed similarly to define the Feynman propagator for general FRW universes.

\subsection{FRW universes}
\label{feynfrw}

For our cosmological model, the Feynman propagator can be defined through a phase-space path integral, taking into account the integration over the lapse $N$ \cite{hennteit}, 
\bena
G(x,m| x',m') &=& \int \mathcal{D}x^\alpha \,\mathcal{D}P_\alpha\,\mathcal{D}m \,\mathcal{D}p_m\,\mathcal{D}N\,\nonumber
\\&& \exp\left(\im\int\limits_{-1/2}^{1/2} dt\;\bigl(\dot{x}^\alpha P_\alpha+\dot{m}p_m\right.
\label{phsppathint}
\\&&\left.-N\left(\frac{P_\alpha P^{\alpha}}{2 m}+ \frac{m}{2} \,  (\kappa \,x^\alpha x_\alpha+ 1)\right)\bigr)\right)\,.\nonumber
\eena
As in the previous example, due to the reparametrization invariance of the theory the parameter time (specified by $t$) between the initial and final configurations is arbitrary, and we choose it to run from $-\half$ to $\half$. In order for the path integral to be well defined, the gauge invariance under time reparametrizations generated by the Hamiltonian constraint must be broken by fixing a specific gauge. One simple gauge fixing, $\dot{N}=0$, can be obtained by introducing a new field $\Pi$ and adding the term $\Pi\dot{N}$ to the action; we refer to Ref.~\cite{halliwdw} for details. Integrating over the field $\Pi$ then reduces the integration over $N$ to an ordinary integral over the total conformal time between the initial and final configurations; we make this explicit by again writing $N$ as $\tau$.  

The remaining path integrals may be computed exactly.  Path integration over $m$ and $p_m$ simply gives a delta function in $m$, as expected since $m$ has trivial dynamics constraining it to be constant. One can then integrate over $P_\alpha$ which yields
\bena
G(x,m| x',m') &&= \delta(m-m')\times\label{eq3a}
\\&&\int  d\tau
\mathcal{D}x\,  \exp\bigl[ \im \frac{m}{2} \int dt\left(\frac{\dot{x}^2}{\tau} -\tau\left(1+\kappa x^2 \right) \right)\bigr] \nonumber
\eena
 corresponding to $M+1$ decoupled harmonic oscillators.

For $\kappa=0$, apart from the overall delta function, this is exactly the expression Eq.~(\ref{e2}). Accordingly, the path integral over $x$ is just that of a free relativistic particle and the $\tau$ integral can be evaluated exactly; the result is
\ben
G^0(x,m| x',m') =\half \delta(m-m') (-\im m)^M (2\pi s)^{\frac{1-M}{2}} H_{\frac{M-1}{2}}^{(2)}(s)\,,
\label{eq4a}
\een 
with $s\equiv m\sqrt{-(x-x')^2-\im \epsilon}$, where $H^{(2)}_\alpha(x)$ is a Hankel function of the second kind (see the appendix). The $-\im\epsilon$ in its argument indicates that the expression is the boundary value of a function which is analytic in the lower half $-(x-x')^2$ plane. As we have emphasized, we have derived this definition from the path integral, and the requirement that the integral over proper time $\tau$ converges. 

In order to understand the more involved case of spatial curvature $\kappa\neq 0$, it is again helpful to recall how Eq.~(\ref{eq4a}) can be obtained from the Wronksian method; for simplicity, let us set $M=0$ and label $x^0\equiv a$ which is our scale factor. Then the Feynman propagator is a solution to
\ben
\left(\frac{1}{2m}\frac{d^2}{da^2}+\frac{m}{2}\right)G^0(a,m| a',m') = -\im\delta(m-m')\delta(a-a')
\een
and can be written in the form
\bena
G^0(a,m|a',m')&=&-2\im m\delta(m-m')\left(\frac{\psi_1(a')\psi_2(a)}{W(\psi_1,\psi_2)}\theta(a-a')\right.\nonumber
\\&&+\left.\frac{\psi_1(a)\psi_2(a')}{W(\psi_1,\psi_2)}\theta(a'-a)\right)
\label{analyticgreens}
\eena
where $W(\psi_1,\psi_2)=\psi_1(a)\psi'_2(a)-\psi'_1(a)\psi_2(a)$ is again the ($a$-independent) Wronskian and $\psi_1(a)$ and $\psi_2(a)$ are two appropriate independent solutions to the homogeneous equation $\left(\frac{1}{2m}\frac{d^2}{da^2}+\frac{m}{2}\right)\psi(a)=0$, found by matching  Eq.~(\ref{analyticgreens}) with the asymptotic behavior of Eq.~(\ref{e2}) at infinity, with $\sigma=(a-a')^2$ (and there are no spacelike directions to be considered). As explained below Eq.~(\ref{fourierint}),  (\ref{e2}) asymptotes to $e^{-\im m a}$ for large positive $a$ at fixed $a'$, and to $e^{\im m a'}$ for large negative $a'$ at fixed $a$; this fixes the modes in (\ref{analyticgreens}) as $\psi_1(a)=e^{\im m a}$ and $\psi_2(a)=e^{-\im m a}$ (up to a normalization that is irrelevant for $G^0$). Thus, one finds
\ben
G^0(a,m|a',m')=\delta(m-m')e^{-\im m|a-a'|}\,,
\een
in agreement with Eq.~(\ref{eq4a}) for $M=0$.

\begin{figure}[htp]
\includegraphics[scale=0.6]{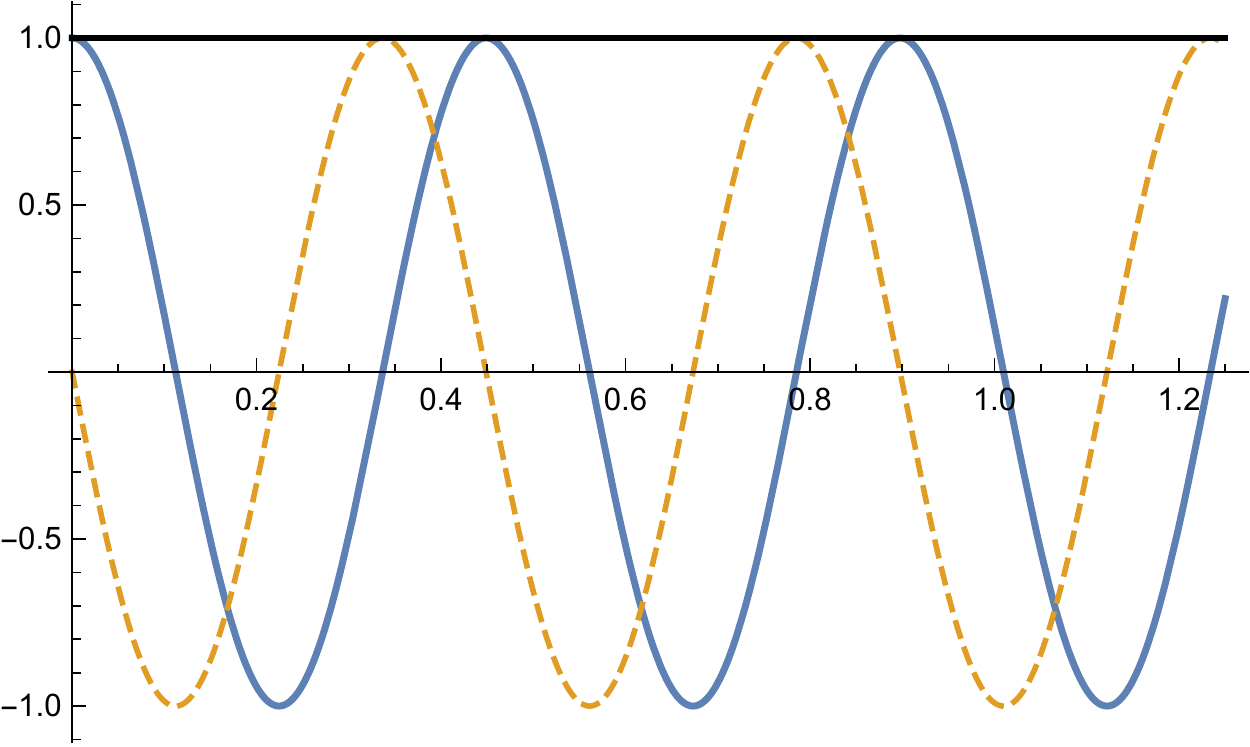}
\caption{Feynman propagator for flat universes as a function of $T$, for $m=7$, showing real part (blue), imaginary part (dashed) and absolute value (black) which is constant here.}
\label{flatpropplot}
\end{figure}

With this in mind, we can now go beyond the simplest flat case, and consider $\kappa\neq 0$, where the dynamics of the Universe corresponds to those of a relativistic oscillator or upside-down oscillator. The path integral over $x$ in Eq.~(\ref{eq3a}) is easily calculated: the classical path which fixes the exponent generalizes to 
 \ben
 x(t)=\frac{x\sin\left(\sqrt{\kappa}\,\tau(t+\half)\right)+x' \sin\left(\sqrt{\kappa}\,\tau(\half-t)\right)}{\sin (\sqrt{\kappa}\, \tau)},
 \label{class1}
 \een
which is unique for all $x$, $x'$ and $\tau$ [where we exclude special cases for which $\sin(\sqrt{\kappa}\,\tau)=0$], and the prefactor is given by the usual regularized functional determinant for the harmonic oscillator, so that
\bena
&&G(x,m| x',m')\nonumber
\\& = &\im\delta(m-m')\int\limits_0^{\infty}  d\tau\,\left(\frac{m \sqrt{\kappa}}{2\im\pi\sin(\sqrt{\kappa}\,\tau)}\right)^{\frac{M+1}{2}}
\label{oscpathint}
\\&&\times\exp\left[\im \,\frac{m}{2} \left(\sqrt{\kappa} \,\frac{(x^2+x'^2)\cos(\sqrt{\kappa}\,\tau)-2x\cdot x'}{ \sin(\sqrt{\kappa}\,\tau)}-\tau\right)\right]\,.\nonumber
\eena
The overall factor of $\im$ arises just as it did for the free relativistic particle, discussed in the previous subsection. As there, we are left with an  ordinary integral over $\tau$, and need to establish the appropriate integration contour. The resulting integral in Eq.~(\ref{oscpathint}) is difficult to do directly but we can use its behavior at large positive $a$ and large negative $a'$ to 
fix the mode functions $\psi_1$ and $\psi_2$  appearing in the Wronskian representation. As a consistency check, we compare the resulting Green's function to a numerical evaluation of the $\tau$ integral in Eq.~(\ref{oscpathint}) along a suitable contour, finding perfect agreement in all cases. In this numerical evaluation, we choose $x=(T,\vec{0})$ and $x'=(-T,\vec{0})$, so that any classical real solution has to pass through the singularity $a=0$ at least once, which is the situation of main relevance for our study. For ease of comparison, we plot the flat case $\kappa=0$, with $M=0$, {\it i.e.}, $G=e^{-2 \im m T} $, with $m=7$, in Fig.~\ref{flatpropplot} [here and in the following we are of course plotting the function multiplying the singular part $\delta(m-m')$]. 

Consider the saddle points in the $\tau$ integral for the curved-space propagator, and the associated steepest descent contour. The condition for the exponent to be stationary with respect to $\tau$ is precisely the Hamiltonian constraint (Friedmann equation) $\tau^{-2} \dot{x}^2 + \left(1+\kappa\, x^2 \right)=0$, with $x(t)$ given by Eq.~(\ref{class1}). Real saddle points of the full functional integral, when they exist, are real solutions of the classical equations of motion, including the constraints. Given such saddle points, one defines the associated $\tau$ integration cycle as the complete extension of the steepest descent contour. If this cycle can be deformed to the real $\tau$-axis while maintaining the convergence of the integral, then the saddle point contribution is relevant to the final result. 

\begin{figure}[htp]
\includegraphics[scale=0.4]{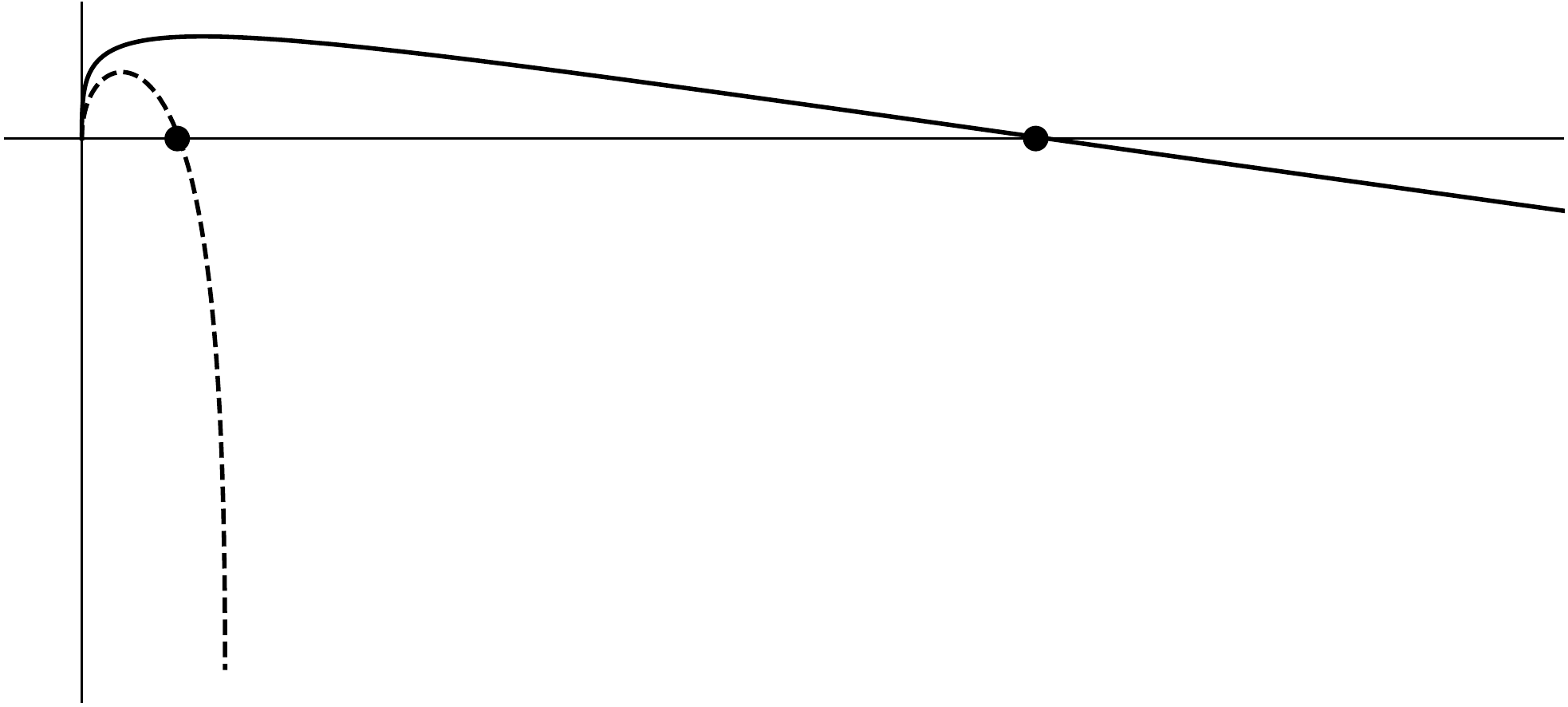}
\caption{Integration contours in the complex $\tau$ plane for $\kappa \leq 0$ (solid line) and for $\kappa >0$ (dashed line), for the case where $x=(T,\vec{0})$ and $x'=(-T,\vec{0})$.}
\label{taucontour}
\end{figure}

We start with the case of negative $\kappa$, where, just as in the flat case, there is always a unique classical solution: for timelike separated $x$ and $x'$, the saddle point in $\tau$ is located on the real axis, and the steepest descent contour is the solid curve in Fig.~\ref{taucontour}. The singular behavior of the integrand at $\tau=0$ [cf. Eq.~(\ref{e5})] then ensures, just as in the argument leading to Eq.~(\ref{e5}), that  Eq.~(\ref{oscpathint}) is a Green's function for the Wheeler-DeWitt equation,
\ben
\left(-\frac{\Box}{2m}+\frac{m}{2}(\kappa x^2+1)\right)G = -\im\delta(m-m')\delta^{M+1}(x-x')\,.
\een

Once more we set $M=0$ for simplicity, and obtain the Green's function from the Wronskian method, with the modes determined from their behavior at large argument. The Wheeler-DeWitt equation, at fixed $m$, is 
\ben
\left(\frac{d^2}{da^2}+m^2(1-\kappa a^2)\right)\psi(a) = 0\,,
\label{webereq}
\een
and is solved by parabolic cylinder functions \cite{abrama}, denoted by $D_\nu(z)$. In order to find the modes that generalize the plane waves $e^{\pm \im m a}$ used in the $\kappa=0$ case, we can again study the asymptotic limits of Eq.~(\ref{oscpathint}) using the saddle-point approximation, finding that we must have (with $\kappa<0$)
\bena
\psi_3(a)&\sim& |a|^{-\half} e^{-\im\frac{m}{2}\sqrt{-\kappa}a^2}\,,\quad a\rightarrow -\infty\,,\nonumber
\\\psi_4(a)&\sim& |a|^{-\half} e^{-\im\frac{m}{2}\sqrt{-\kappa}a^2}\,,\quad a\rightarrow +\infty\,,
\label{asymptotics}
\eena
for the  mode functions $\psi_3$ and $\psi_4$ appearing in the analog of Eq.~(\ref{analyticgreens}). This asymptotic behavior is also consistent with the requirement that, as Eq.~(\ref{oscpathint}) is invariant under $x\rightarrow -x$ and $x'\rightarrow -x'$, $\psi_3$ and $\psi_4$ must satisfy
\ben
\psi_3(-a)=\psi_4(a)\,.
\een
Two independent solutions to Eq.~(\ref{webereq}) are given by
\bena
\psi(a)&=&D_{\im\frac{m}{2\sqrt{-\kappa}}-\half}((1-\im)\sqrt{m}\,(-\kappa)^{1/4}a)
\eena
and its complex conjugate, which asymptotically become pure negative and positive frequency modes as $a\rightarrow\infty$ but are a mixture as $a\rightarrow-\infty$. We therefore set
\ben
\psi_4(a)=D_{-\im\frac{m}{2\sqrt{-\kappa}}-\half}((1+\im)\sqrt{m}\,(-\kappa)^{1/4}a)
\een
and $\psi_3(a)=\psi_4(-a)$. By computing their Wronskian we obtain the Green's function
\bena
G(a,m|a',m')\,&&=\frac{\sqrt{\im m}}{\sqrt{\pi}(-\kappa)^{1/4}}\,\Gamma\left(\half+\frac{\im m}{2\sqrt{-\kappa}}\right)\delta(m-m')\nonumber
\\&&\times\left(\psi_3(a')\psi_4(a)\theta(a-a')+(a\leftrightarrow a')\right)\,.
\label{paragreens}
\eena

\begin{figure}[htp]
\includegraphics[scale=0.6]{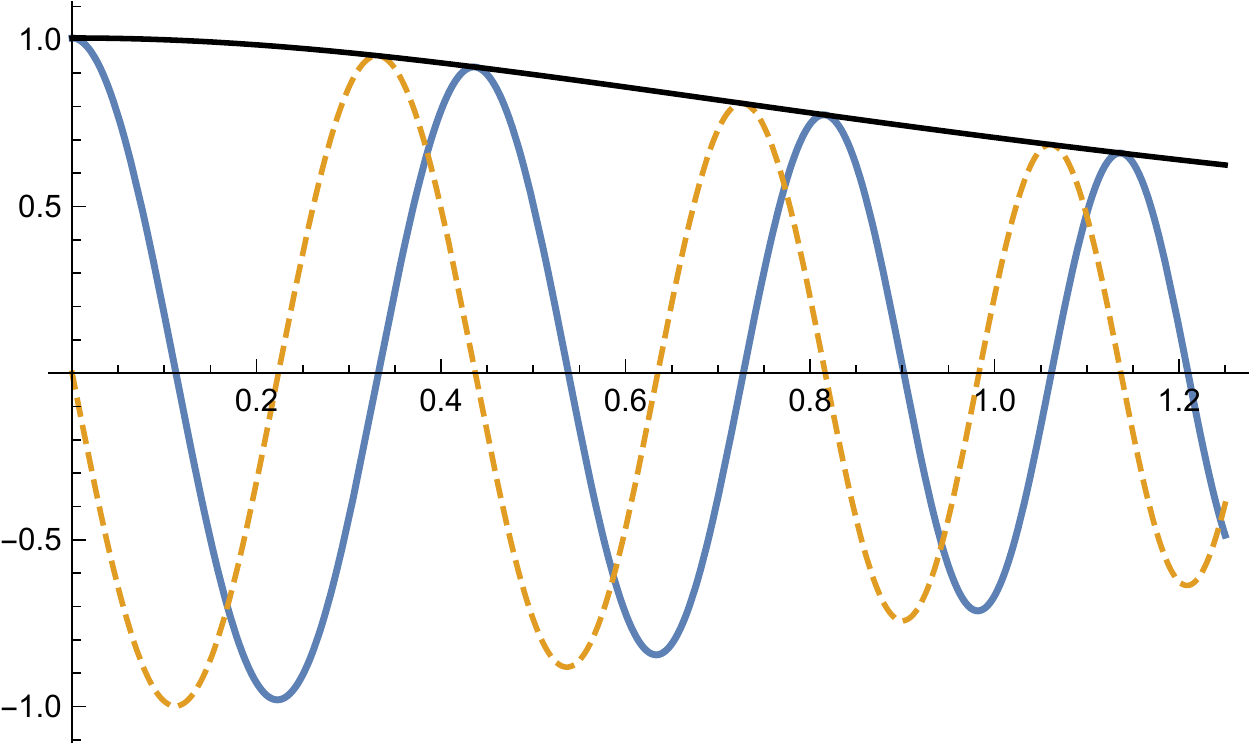}
\caption{Feynman propagator for open universes as a function of $T$, for $m=7$ and $\kappa=-1$, showing real part (blue), imaginary part (dashed) and absolute value (black).}
\label{openprop}
\end{figure}

As we have said, this result can also be obtained from numerical evaluation of Eq.~(\ref{oscpathint}). Again we choose $x=(T,\vec{0})$ and $x'=(-T,\vec{0})$ and also fix $\kappa=-1$, $M=0$ and $m=7$. The resulting function of $T$ is plotted in Fig.~\ref{openprop}. Notice that the resulting propagator resembles the flat-space ($\kappa=0$) expression  for small $T$, and the effects of spatial curvature become relevant only at scales $|x|\sim\frac{1}{\sqrt{|\kappa|}}$. 

For positive $\kappa$, the behavior is rather different from $\kappa\le 0$ in that for positive $\kappa$ the real, classical solutions are periodic in $\tau$; for given $x$ and $x'$, when one classical solution exists there will be an infinite number, and in general they should all contribute to the propagator. Again for consistency with the $\kappa\rightarrow 0$ limit, we can choose the $\tau$ integration contour such that it only picks out the simplest saddle point, where the classical solution interpolating between $x$ and $x'$ has no turning points. The corresponding saddle point and steepest descent contour, indicated by the dashed curve in Fig.~\ref{taucontour}, goes over to the unique $\kappa=0$ saddle point and steepest descent contour in Fig.~\ref{conts} in the flat limit $\kappa\rightarrow 0$.

\begin{figure}[htp]
\includegraphics[scale=0.6]{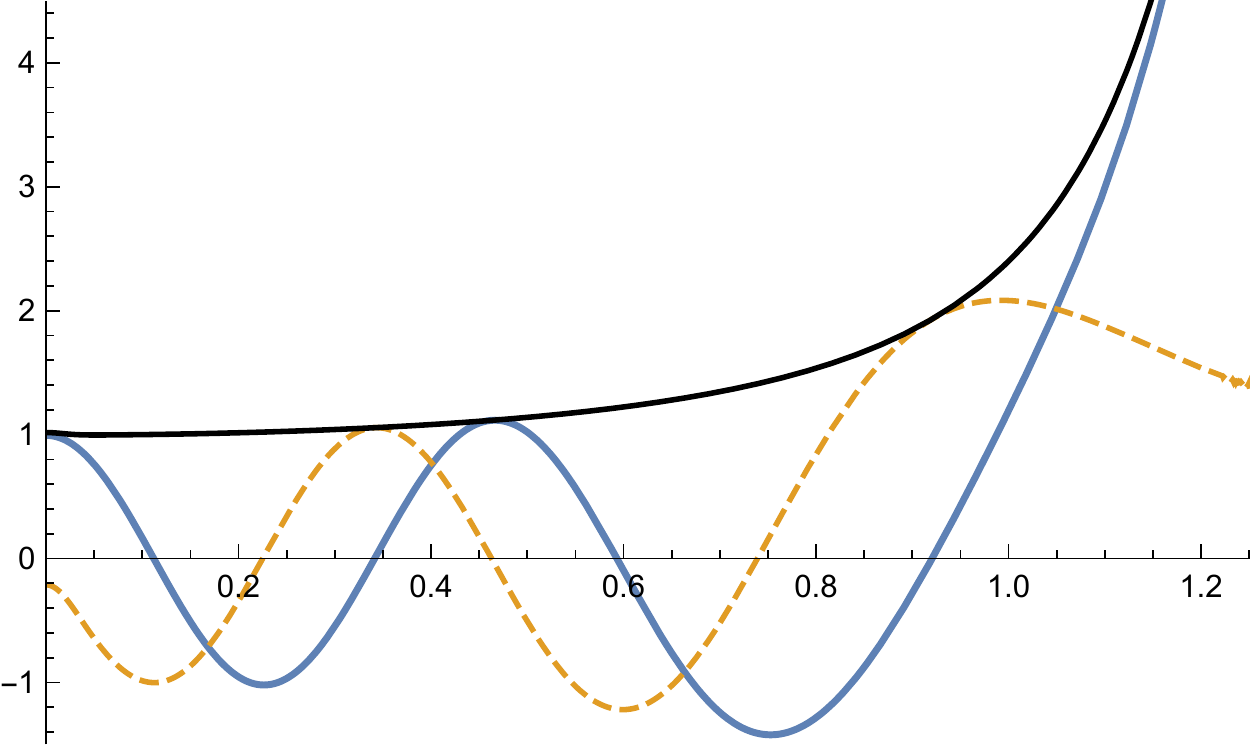}
\caption{Feynman propagator for closed universes as a function of $T$, for $m=7$ and $\kappa=1$, showing real part (blue), imaginary part (dashed) and absolute value (black). As discussed in the text, for $T\geq1$ the semiclassical interpretation fails.}
\label{propfig}
\end{figure}

Another issue is that for large timelike $x$ and $x'$ there is {\it no} real classical solution at all; for large timelike arguments, the saddle point for $\tau$ becomes imaginary and, as explained above, we choose the one on the negative imaginary axis. By the same saddle-point approximation as above, one then determines the asymptotic behavior of the relevant mode functions $\psi_5$ and $\psi_6$,
\bena
\psi_5(a)&\sim& |a|^{-\frac{m}{2\sqrt{\kappa}}-\half} e^{+\frac{m}{2}\sqrt{\kappa}a^2}\,,\quad a\rightarrow -\infty\,,
\\\psi_6(a)&\sim& |a|^{-\frac{m}{2\sqrt{\kappa}}-\half} e^{+\frac{m}{2}\sqrt{\kappa}a^2}\,,\quad a\rightarrow +\infty\,,
\eena
again consistent with $\psi_5(a)=\psi_6(-a)$. These results, as well as the form of the propagator, can in fact be obtained by replacing $\sqrt{-\kappa}\rightarrow\im\kappa$ in the expressions for the open case. [The factor $|a|^{-m/(2\sqrt{\kappa})}$ was dropped in the expressions above since it was a subleading oscillatory factor $|a|^{-\im m/(2\sqrt{-\kappa})}$.]

It is straightforward to obtain a complete analytic expression for the $\kappa>0$ propagator from parabolic cylinder functions by solving the homogeneous equation (\ref{webereq}). Two independent solutions with $\kappa>0$ are
\bena
\psi_5(a)&=&D_{-\frac{m}{2\sqrt{\kappa}}-\half}(-\im\sqrt{2m}\,\kappa^{1/4}a)\,,\nonumber
\\\psi_6(a)&=&D_{-\frac{m}{2\sqrt{\kappa}}-\half}(\im\sqrt{2m}\,\kappa^{1/4}a)\,.
\label{parabolic}
\eena
These are again complex conjugates of each other, and here also already satisfy $\psi_5(a)=\psi_6(-a)$, unlike for the open case $\kappa<0$. On the other hand, they are not asymptotic positive or negative frequency modes, but blow up exponentially both at positive and at negative infinity. At small $a$, up to corrections that vanish as $\kappa\rightarrow 0$ they reduce to plane waves $e^{\pm\im m a}$. From Eq.~(\ref{parabolic}), the Wronskian method gives the Green's function
\bena
G(a,m|a',m')\,&&=\frac{\sqrt{m}}{\sqrt{\pi}\,\kappa^{1/4}}\Gamma\left(\half+\frac{m}{2\sqrt{\kappa}}\right)\,\delta(m-m')\times\nonumber
\\&&\left(\psi_5(a')\psi_6(a)\theta(a-a')+(a\leftrightarrow a')\right)\,.
\label{greensclosed}
\eena
As before, we have verified that this expression agrees with the result of a numerical integration of the $\tau$ integral along the chosen contour. With $x=(T,\vec{0})$, $x'=(-T,\vec{0})$, as well as $\kappa=1$, $M=0$ and $m=7$, the resulting function of $T$ is plotted in Fig.~\ref{propfig}. Again, it reduces to the $\kappa=0$ expression $e^{-2\im m T}$ for small $T$. 

The integration contour in Fig.~\ref{taucontour} is chosen in such a way that its main contribution, for small enough $T$, comes from the lowest positive real saddle point in $\tau$, corresponding to the real classical solution that takes the smallest amount of proper time. As the arguments of the Feynman propagator approach $x=(1/\sqrt{\kappa},\vec{0})$ and $x'=(-1/\sqrt{\kappa},\vec{0})$, this saddle point moves towards $\tau=\frac{\pi}{\sqrt{\kappa}}$ where it eventually merges with another saddle point approaching $\tau=\frac{\pi}{\sqrt{\kappa}}$ from above, corresponding to two classical solutions that become indistinguishable in this limit. Our choice of integration contour then becomes ambiguous and is no longer defined by consistency with the $\kappa\rightarrow 0$ limit. As we extend the arguments to $|T|> 1$, where there is no longer a real solution, these two saddle points separate again and start moving up and down in the imaginary direction. This is akin to the situation for spacelike separations for the relativistic particle, and means that our saddle point needs to be replaced by a saddle point on the line $\frac{\pi}{\sqrt{\kappa}}-\im\bR$ parallel to the negative imaginary $\tau$-axis. For the purposes of this paper, we are mainly interested in studying the propagator with arguments for which there is a classical solution, so that a semiclassical picture of the propagator as given by these solutions is meaningful. 

The exponential blowup of the Feynman propagator for large $T$ follows from the asymptotic behavior of the integral (\ref{oscpathint}) for large timelike $x$ and $x'$. The corresponding mode functions increase exponentially for $|T|> 1$ when there are no classical solutions, as can be verified explicitly from the asymptotics of the parabolic cylinder functions (\ref{parabolic}). The Feynman propagator is hence pathological for large timelike separations, and does not define a suitable two-point function on the entire superspace, because positive curvature forces classical solutions to recollapse. An asymptotic description in terms of well-defined states that can be used to formulate a quantum theory of expanding universes does not exist for positive spatial curvature, and this case does not consistently describe the type of quantum bounce we are interested in. Of course, this situation could be altered by the inclusion of a positive cosmological constant, as we mention later. 

For $\kappa<0$, classical solutions with pure radiation are well behaved, expanding to infinite volume in the future and past and leading to a well-behaved Feynman propagator, Eq.~(\ref{asymptotics}). However, as we mentioned in Sec.~\ref{friedbounce}, when conformally coupled scalars are introduced ($M\ge 1$), the general solution
\bena
x^{\alpha}(t)&=&\frac{x_1^\alpha}{\sqrt{-\kappa}} \exp\left(\sqrt{-\kappa}\int\limits_0^t dt'\,N(t')\right)\nonumber
\\&&+\frac{x_2^\alpha}{\sqrt{-\kappa}} \exp\left(-\sqrt{-\kappa}\int\limits_0^t dt'\,N(t')\right)\,,
\eena
with $x_1\cdot x_2=\frac{1}{4}$, has both spacelike antigravity and timelike gravity solutions: choosing spacelike $x_1$ and $x_2$ that satisfy the constraint, one finds a universe that comes in from one antigravity direction, turns around before entering gravity and then disappears into a different (or the same) antigravity direction. Such spacelike solutions are far enough in the antigravity region that the curvature term dominates over the positive mass in the potential, $m^2(1+\kappa x^2)<0$, leading to their acceleration towards spacelike infinity. Even though there are no real classical solutions that connect incoming gravity solutions to these  far antigravity regions, and starting in a gravity region one is guaranteed to asymptote into the other gravity region, quantum mechanically one expects the spacelike solutions to determine the behavior of the Feynman propagator for spacelike separations. This is indeed confirmed by finding the saddle-point approximation to Eq.~(\ref{oscpathint}) for large spacelike separations, {\it e.g.}, for $x^2\rightarrow \infty$ at fixed $x'$,
\ben
G(x,m|x',m')\sim |x|^{-1/2}e^{\im\frac{m}{2}\sqrt{-\kappa}x^2}\,.
\een
The propagator becomes oscillatory at spacelike separations, so that a given initial state, {\it e.g.}, a wave packet centered around an initial state in the gravity region $a<0$, is propagated to large spacelike distances into the antigravity region. In this sense, the quantum theory is even more pathological for open than for closed universes, where one finds, again for $x^2\rightarrow \infty$ at fixed $x'$,
\ben
G(x,m|x',m')\sim |x|^{-\frac{m}{2\sqrt{\kappa}}-\half} e^{-\frac{m}{2}\sqrt{\kappa}x^2}\,,
\een
{\it i.e.}, exponential falloff for large spacelike separations. This is because for $\kappa>0$, both timelike and spacelike classical solutions are bounded due to the potential, and never reach (timelike or spacelike) infinity. Neither $\kappa<0$ nor $\kappa>0$ can lead to a viable perfect bounce picture in terms of a transition between incoming and outgoing gravity states with $a\rightarrow\pm\infty$, while the $\kappa=0$ case leads directly to a perfect bounce. We conclude that, at least for the theories considered here, only flat FRW universes lead to a consistent quantum theory, able to account for an expanding universe. The inclusion of a positive cosmological constant could rescue positively curved universes from this conclusion, provided the curvature is too small to cause a recollapse. Nevertheless, the quantum pathology we have identified for 
negatively curved FRW universes is intriguing, because it raises the possibility that the observed (nearly flat) Universe lives on the corresponding critical boundary. This would be the case, for example, if we could identify the correct quantum measure on the space of closed universes, with sufficiently large cosmological constant to prevent recollapse, and if this measure favored  the flat case. We explore this possibility in future work. For the remainder of this paper, however, we ignore spatial curvature.

\section{Adding anisotropies and free scalar fields}
\label{anisosec}

We now extend the treatment to anisotropic cosmologies, choosing the simplest form of anisotropies, the Bianchi I model: we still require the metric to be spatially homogeneous, with an Abelian group of isometries acting on constant time hypersurfaces, but no longer impose isotropy. The most convenient parametrization of such a metric employs Misner variables \cite{misner},
\bena
ds^2&=&A^2(t)\left(-N^2(t)dt^2+e^{2\lambda_1(t)+2\sqrt{3}\lambda_2(t)}dx_1^2\right.\nonumber
\\&&\left.+e^{2\lambda_1(t)-2\sqrt{3}\lambda_2(t)}dx_2^2+e^{-4\lambda_1(t)}dx_3^2\right)\,.
\label{bianchimetric}
\eena
The Ricci tensor, and hence the Einstein tensor, for this metric are diagonal, which by the Einstein equations forbids any anisotropy in the fluid, manifest in a velocity $u^i$. We hence continue to assume that the fluid moves with the cosmological flow, $J^\mu\propto\delta^\mu_0$. 

We can then again exploit the conformal freedom to set $A(t)$ to 1. The Ricci scalar of (\ref{bianchimetric}) is then 
\ben
R=6\frac{\dot\lambda_1^2+\dot\lambda_2^2}{N^2}\,,
\een
giving the correct canonical normalization of the anisotropy variables $\lambda_i$ in the symmetry-reduced action,
\ben
S=V_0\int dt\,\bigl[\frac{\dot{\vec{\chi}}^2-\dot{\phi}^2}{2 N}+\frac{\dot\lambda_1^2+\dot\lambda_2^2}{2N}(\phi^2-\vec\chi^2) -N\rho-\tilde\varphi \dot{n}  \bigr]\,.
\label{symraction}
\een
In terms of the variables $x^\alpha$ and $m$ defined in Eq.~(\ref{newvariables}), this action reads
\ben
S= \int dt \bigl[\frac{m}{2}\left(\frac{1}{N}\left(\dot{x}^2-x^2(\dot\lambda_1^2+\dot\lambda_2^2)\right)-N\right)-\varphi\dot{m}\bigr]\,.
\label{bianchiaction}
\een
As for the flat FRW universe which the Bianchi I universe generalizes, this is the action of a free massive relativistic particle. However, here the particle is not moving through a flat Minkowski spacetime but through a curved superspace, with metric
\ben
ds^2=\eta_{\alpha\beta}dx^\alpha dx^\beta-x^2(d\lambda_1^2+d\lambda_2^2)
\een
or, if we again use the parametrization $x^\alpha=a v^\alpha(\nu^i)$ with $v^2=-1$ to separate the scale factor $a$, introducing coordinates $\nu^i$ on the hyperboloid $H^M$,
\ben
ds^2=-da^2+a^2\,g_{H^M}+a^2(d\lambda_1^2+d\lambda_2^2)
\een
where $g_{H^M}$ is a constant negative curvature metric on $H^M$, as in Sec.~\ref{friedbounce}. The geometry of superspace at fixed $a$ corresponds to the maximally symmetric space $H^M\times\bR^2$; in the absence of anisotropies, the last term would vanish and one would simply use Milne coordinates for flat Minkowski spacetime.

It is well known that the dynamics of anisotropies in the Bianchi I model are equivalent to those of minimally coupled free scalar fields in this background. To see this, we momentarily switch to Einstein gauge, in which the metric is given by Eq.~(\ref{bianchimetric}) with a general $A(t)$. The action of a free scalar field in this background is
\ben
-\half\int d^4 x\,\sqrt{-g}\,(\partial\tau)^2 = V_0\int dt\;A^2\frac{\dot{\tau}^2}{2N}
\label{scalaraction}
\een
(identical to the action of a free scalar in a flat FRW universe; a homogeneous scalar field does not feel the anisotropies). To see that this reduces to the kinetic terms for the anisotropies $\lambda_i$ in Eq.~(\ref{symraction}), we note that since the scale factor (\ref{scalefactor}) is Weyl invariant, one can express it both in Weyl and Einstein gauge,
\ben
a^2=\frac{1}{2\rho}(\phi^2-\vec\chi^2)\big|_{{\rm Weyl}}=\frac{3A^2}{8\rho_0\pi G}\big|_{{\rm Einstein}}
\een
with $\rho=\rho_0 A^{-4}$ in Einstein gauge (in Weyl gauge, $\rho=\rho_0$ is constant). To change gauges, one hence has to replace
\ben
A^2\big|_{{\rm Einstein}} \rightarrow \frac{4\pi G}{3}(\phi^2-\vec\chi^2)\big|_{{\rm Weyl}}\,;
\een
the factor $\frac{4\pi G}{3}$ can be absorbed in the normalization of the scalar fields,
\ben
\lambda:=\sqrt{\frac{4\pi G}{3}}\tau
\label{normalize}
\een
(the anisotropy variables are dimensionless while a scalar field has dimensions of mass), showing the equivalence. Hence, we obtain a simple generalization of the theory we have discussed by also adding an arbitrary number of minimally coupled free scalar fields, which can represent either physical scalar fields or anisotropy degrees of freedom of the Bianchi I model. One has to lift the free scalar fields to a Weyl-invariant theory by
\ben
-\half\int d^4 x\,\sqrt{-g}\,(\partial\tau)^2 \rightarrow -\half\int d^4 x\,\sqrt{-g}\,(\phi^2-\vec\chi^2)(\partial\lambda)^2
\label{scalarlift}
\een
where $\lambda$ is again dimensionless and conformally invariant. In Einstein gauge, the right-hand side of Eq.~(\ref{scalarlift}) clearly reduces to the right-hand side of Eq.~(\ref{scalaraction}). Going back to Weyl gauge, the total action is then
\ben
S= \int dt \bigl[\frac{m}{2}\left(\frac{1}{N}\left(\dot{x}^2-x^2\sum_{i=1}^K\dot\lambda_i^2\right)-N\right)-\varphi\dot{m}\bigr]\,,
\label{generalaction}
\een
which is a simple generalization of Eq.~(\ref{bianchiaction}). The $K$ variables $\lambda_i$, $i=1,\ldots,K$, can now correspond to anisotropy variables or minimally coupled scalar fields with the unusual normalization (\ref{normalize}). Equation (\ref{generalaction}) is now the action of a particle moving in an $(M+K+1)$-dimensional superspace with curved metric
\bena
ds^2&=&\eta_{\alpha\beta}dx^\alpha dx^\beta-x^2\sum_{i=1}^K d\lambda_i^2\nonumber
\\&=&-da^2+a^2\,g_{H^M}+a^2\sum_{i=1}^K d\lambda_i^2\,.
\label{supermetric}
\eena
By an elementary generalization of the procedure described in Sec.~\ref{canonical}, Eq.~(\ref{generalaction}) gives a Hamiltonian constraint
\ben
C:=\frac{1}{2m}g^{\mu\nu}(x,\lambda)p_\mu p_\nu + \frac{m}{2}\approx 0
\label{generalconstraint}
\een
where $g^{\mu\nu}$ is the inverse metric on superspace, and $p_\mu$ includes conjugate momenta for both the variables $x^\alpha$ and the new degrees of freedom $\lambda_i$; more concretely,
\ben
C=\frac{1}{2m}\left(-p_a^2+\frac{1}{a^2}g^{ij}_{H^M}(\nu)\zeta_i\zeta_j +\frac{1}{a^2}\delta^{ij}k_i k_j\right)+\frac{m}{2} 
\label{constraint}
\een
in terms of the momentum $p_a$ conjugate to $a$, momenta $\zeta_i$ conjugate to the conformally coupled scalar field variables $\nu^i$ living on $H^M$, and momenta $k_i$ conjugate to the free scalar fields and anisotropy variables.

When quantizing this Hamiltonian constraint in order to obtain the Wheeler-DeWitt equation, there is now an ambiguity, the well-known quantization ambiguity for a particle moving on a curved manifold \cite{dewitt2}: if the Ricci scalar for the superspace metric (\ref{supermetric}) is $\mathcal{R}$, the general expression for the quantum Hamiltonian is
\ben
H=\frac{1}{2m}\left(-\Box+\xi\mathcal{R}\right)+\frac{m}{2}
\label{quantham}
\een
where $\Box$ is the Laplace-Beltrami operator for the curved metric (corresponding to the operator ordering that ensures that the Hamiltonian is independent of the choice of coordinates on superspace) and $\xi$ is, in general, a free parameter. Halliwell \cite{halliwdw} has given the following strong argument for fixing $\xi$: the (classical) Hamiltonian of minisuperspace models is really $\mathcal{H}=N\,C$ (see Sec.~\ref{friedbounce}) where the lapse function $N$ is arbitrary, and in particular can be redefined arbitrarily, $N\rightarrow\Omega^{-2}\tilde{N}$, where $\Omega(x,\lambda)$ is any function on minisuperspace. Under such a redefinition, the constraint (\ref{generalconstraint}) becomes
\ben
\tilde{C}:=\frac{1}{2m}\tilde{g}^{\mu\nu}(x,\lambda)p_\mu p_\nu + \frac{\tilde{m}(x,\lambda)}{2}\approx 0
\een
with $\tilde{g}^{\mu\nu}=\Omega^{-2}g^{\mu\nu}$ and $\tilde{m}(x,\lambda)=\Omega^{-2}m$, leading by the same argument as above to a quantum Hamiltonian
\ben
\tilde{H}=\frac{1}{2m}\left(-\tilde{\Box}+\xi\tilde{\mathcal{R}}\right)+\frac{\tilde{m}(x,\lambda)}{2}\,,
\label{qhricc}
\een
where now $\tilde{\Box}$ and $\tilde{\mathcal{R}}$ are the Laplace-Beltrami operator and Ricci scalar for the conformally rescaled metric $\tilde{g}$ on superspace. Halliwell now asks that, since redefining the lapse is always possible classically, the solutions $\Psi$ and $\tilde\Psi$ to $H\Psi=0$ and $\tilde{H}\tilde\Psi=0$ be related by a conformal transformation, $\tilde\Psi=\Omega^\gamma\Psi$ for some $\gamma$, and finds that this is only possible if one fixes $\xi$ to be the conformal coupling,
\ben
\xi=\frac{M+K-1}{4(M+K)}
\label{xivalue}
\een
(recall that the dimension of our superspace manifold is $M+K+1$; Ref.~\cite{halliwdw} gives an overall minus sign for $\xi$, presumably due to a different sign convention for the Ricci curvature). Demanding covariance under field redefinitions of the lapse function hence fixes $\xi$ uniquely. Special cases are $M+K=0$ where there is no conformal coupling that can restore covariance under lapse redefinitions, and $M+K=1$ where the Laplace-Beltrami operator is conformally covariant and $\xi=0$.

The Wheeler-DeWitt equation for $\Psi=\Psi(a,\nu,\lambda,m)$, corresponding to the classical constraint (\ref{constraint}), then becomes
\ben
\left(\frac{\partial^2}{\partial a^2}+\frac{M+K}{a}\frac{\partial}{\partial a}-\frac{1}{a^2}\Delta_{H^M\times\bR^K}+\xi\mathcal{R}+m^2\right)\Psi=0
\een
with
\ben
\mathcal{R}=\frac{K(2M+K-1)}{a^2}\,.
\een
As in Sec.~\ref{friedbounce}, one can simplify the Wheeler-DeWitt equation by Fourier transforming on $H^M\times\bR^K$ from $\nu$ and $\lambda$ to the momenta $\zeta$ and $k$. One then has
\ben
\left(\frac{\partial^2}{\partial a^2}+\frac{M+K}{a}\frac{\partial}{\partial a}-\frac{c}{a^2}+m^2\right)\Psi(a,\zeta,k,m)=0
\label{wdwbianchi}
\een
with
\ben
c=-\frac{1}{4}(M-1)^2+\frac{1}{4}\delta_{M,0}-\vec\zeta^2-\vec{k}^2-\xi K (2M+K-1)\,,
\label{constant}
\een
where we have explicitly included the case $M=0$ through the Kronecker delta. This is precisely the same functional form as the Wheeler-DeWitt equation for FRW universes, Eq.~(\ref{Wdwfrw}), and so the extension of our formalism from FRW symmetry to the Bianchi I model and the inclusion of minimally coupled scalars are completely straightforward. The constant $c$ now gets contributions from the eigenvalues of the Laplacian on $H^M\times\bR^K$, as well as from the curvature on superspace.

We can now obtain the general solution to Eq.~(\ref{wdwbianchi}) in the usual way, by setting
\ben
\Psi(a,\zeta,k,m)=a^{-(M+K)/2}\chi(a,\zeta,k,m)
\label{psichi}
\een
to eliminate the first derivative. $\chi$ then satisfies the differential equation
\ben
\left(\frac{\partial^2}{\partial a^2}-\frac{c'}{a^2}+m^2\right)\chi(a,\zeta,k,m)=0
\label{schrodinger}
\een
where $c'\equiv c+\frac{1}{4}(M+K)(M+K-2)$, which has two independent solutions in terms of Bessel functions of the first and second kind,
\ben
\chi_1=\sqrt{a}\,J_{\half\sqrt{1+4c'}}(ma)\,,\;\chi_2=\sqrt{a}\,Y_{\half\sqrt{1+4c'}}(ma)\,.
\een
A more convenient basis is given by the Hankel functions $H^{(1,2)}$ of the first and second kind (which are just linear combinations of the Bessel functions), so that two linearly independent solutions of Eq.~(\ref{wdwbianchi}), for fixed $\zeta$, $k$ and $m$, are finally given by
\ben
\psi_{+,-}(a)=a^{-(M+K-1)/2}H_{\half\sqrt{1+4c'}}^{(2,1)}(ma)\,.
\label{plusminus}
\een
As indicated by the subscript $+,-$, these functions represent positive- and negative-frequency modes for the Wheeler-DeWitt equation. Indeed, when extended to negative $a$ through the analytic continuation (see, {\it e.g.}, Ref.~\cite{tolleyturok}) $H_\nu^{(2)}(-z)=-e^{\im\pi\nu}H_\nu^{(1)}(z)$ and $H_\nu^{(1)}(-z)=-e^{-\im\pi\nu}H_\nu^{(2)}(z)$, $\psi_{+,-}$ have the interesting property of corresponding to pure positive and pure negative frequency, respectively, both at positive and negative infinite $a$,
\bena
\psi_{+}(a)&\sim& a^{-(M+K)/2}e^{-\im m a}\,,\quad a\rightarrow\pm\infty\,,\nonumber
\\\psi_{-}(a)&\sim& a^{-(M+K)/2}e^{+\im m a}\,,\quad a\rightarrow\pm\infty\,.
\label{hankelasymptotics}
\eena
That is, for the Wheeler-DeWitt equation (\ref{wdwbianchi}) one finds that an incoming positive-frequency mode simply continues to an outgoing positive-frequency mode, with the potential at $a=0$ not even leading to a phase shift. This complete invisibility of the $1/a^2$ potential is a direct consequence of the symmetry of Eq.~(\ref{wdwbianchi}) under $a\rightarrow \lambda a$ and $m\rightarrow \lambda^{-1}m$, which forbids any phase shift. These special properties of a $1/x^2$ potential, and its invisibility in a scattering process, are well known in quantum mechanics. In the context of our perfect bounce scenario, they imply that the Universe can go through the singularity $a=0$ without any noticeable impact on its evolution, when viewed asymptotically. This is already true classically, where the classical solutions bounce without any net time delay or advance: the classical Hamiltonian is equal to the constraint (\ref{constraint}) times a lapse function,
\ben
\mathcal{H}=N\left(\frac{1}{2m}\left(-p_a^2+\frac{1}{a^2}g^{ij}_{H^M}(\nu)\zeta_i\zeta_j +\frac{1}{a^2}\vec{k}^2\right)+\frac{m}{2}\right)
\een
The terms multiplying $1/a^2$ are again conserved and can be replaced by a constant, $-c'$ with $c'<0$; classically the effect of anisotropies and momenta in the scalar fields always leads to an attractive potential for $a$, centered on the singularity at $a=0$. The classical solutions to the equations of motion including the constraint are then
\ben
a^2=\frac{c'}{m^2}+N^2(t-t_0)^2=\frac{c'}{m^2}+(\tau-\tau_0)^2
\label{solutions}
\een
in terms of proper time $\tau=N t$. These solutions are singular at $a=0$ and perform an excursion into the antigravity region of imaginary $a$, just as the generic flat FRW solutions described in Sec.~\ref{friedbounce} which would be of the exact same form. The attractive potential at $a=0$ speeds up the trajectory as it heads toward the singularity, but this time advance is canceled by the additional time it takes to cross antigravity. Indeed, both at large positive and negative $a$ we have simply $a(\tau)\approx(\tau-\tau_0)$. 

In the quantum theory, ordering ambiguities in the Hamiltonian constraint can alter the coefficient of the $1/a^2$ potential, making it repulsive in some cases. Indeed, the relevant coefficient of the potential is the one appearing in Eq.~(\ref{schrodinger}),
\bena
c'&=&-\frac{1}{4}(M-1)^2+\frac{1}{4}\delta_{M,0}-\vec\zeta^2-\vec{k}^2\nonumber
\\&&+\frac{M^2(M-2)+K(M^2-M-1)}{4(M+K)}
\label{cprime}
\eena
if we use the value (\ref{xivalue}) for $\xi$, that is, we fix the ordering ambiguities by demanding coordinate covariance on superspace and covariance under redefinitions of the lapse function, giving a purely quantum contribution in the second line of Eq.~(\ref{cprime}). If we ignore the trivial case $M=K=0$, we can rewrite Eq.~(\ref{cprime}) as
\ben
c' = \begin{cases}
-\vec{\zeta}^2-\frac{1}{4} & M\ge 1\,,\;K=0\,,
\cr -\vec{k}^2-\frac{1}{4} & K\ge 1\,,\;M=0\,,
\cr -\zeta^2-\vec{k}^2-\frac{1}{4} &  K \ge 1\,,\;M=1\,,
\cr -\vec\zeta^2-\vec{k}^2-\frac{1}{4}+\frac{(M-1)K}{4(M+K)} & K \ge 1\,,\;M> 1\,.
\label{cpp}
\end{cases}
\een
This is an intriguing result. The contributions coming from anisotropy or scalar field momenta are both negative. The numerical term is fixed by covariance.  The first line corresponds to the situation of Sec.~\ref{friedbounce}, where no anisotropies or minimally coupled scalars are present [from Eq.~(\ref{Wdwfrw}), removing the first derivative term changes $c$ in Eq.~(\ref{cpminimal}) to $c'$ given here]. The similarity of the first three lines is not a coincidence; for $M\le 1$, the superspace metric (\ref{supermetric}) is conformally flat. As we have imposed conformal coupling to the Ricci scalar on superspace in Eqs.~(\ref{qhricc}) and (\ref{xivalue}), the dynamics must be equivalent to the flat superspace case of Sec.~\ref{friedbounce}.

The value $c'=-\frac{1}{4}$ is well known as a critical value in the quantum mechanics of an inverse square potential. If $c'\geq -\frac{1}{4}$, the negative classical potential is outweighed by the kinetic energy due to the Heisenberg uncertainty principle, rendering the energy spectrum strictly positive. There are various infrared regularized versions of the theory in which the spectrum is made discrete by including a positive harmonic potential~\cite{parisi}, with $a$ taken either on the infinite line, the half-line $a>0$, or by imposing periodicity in $a$, in which case the model becomes the Calogero-Sutherland model (see, {\it e.g.}, Ref.~\cite{lapointe}). These are well-defined, exactly solvable models which exhibit, among other interesting phenomena, anomalous dimensions in the operator product expansion~\cite{parisi}.

If, however,  $c'<-\frac{1}{4}$, any finite energy wave function has an infinite number of oscillations on the way to $a=0$. In quantum mechanics, standard arguments then imply an infinite number of lower energy states, and hence a spectrum which is unbounded below. It has been claimed that the theory is nevertheless renormalizable, although the renormalization group displays a limit cycle~\cite{braaten}. (There is a large literature on inverse square potentials in quantum mechanics, and even some experimental tests. See, {\it e.g.}, Ref.~\cite{alhaidari} for a recent discussion and further references.)

At the minisuperspace level discussed here, negative energy states are irrelevant because we are only interested in solutions of the Wheeler-DeWitt equation with positive energy, defined by $m^2$. However, when we include interactions with other modes, such as the inhomogeneous modes of gravitons or scalars, then for $c'<-\frac{1}{4}$ it is possible that the negative energy states for the scale factor $a$ would become excited, potentially signifying strong backreaction as the Universe passes through the quantum bounce. The problem may be avoided in two ways.  For $M=0$ or $M=1$ one can restrict consideration to background cosmologies for which the zero-mode momenta of the anisotropy and scalar fields are strictly 0, in which case the quantum mechanics for $a$ lies on the critical boundary where it (just) makes sense. Or, one can include additional conformally coupled scalars, taking $M>1$ so that, from the last line of Eq.~(\ref{cpp}), the quantum mechanics of $a$ is well defined for a range of classical anisotropy and scalar field momenta. For $K=2$ ({\em i.e.}, only anisotropies but no minimally coupled scalars) and $M>4$, the numerical contribution can be large enough to make the potential repulsive at small momenta. If we consider classical solutions with this (order $\hbar$ squared) potential, an isotropic universe with no scalar momenta would bounce off the repulsive potential and avoid the singularity altogether. Quantum mechanically, however, if we extend the range of $a$ to negative values, then $a$ tunnels through the barrier in a process which may be described with complex classical solutions, as we explained in Ref.~\cite{letter}.

The conclusion is that when anisotropy and scalar field degrees of freedom are included, then for small numbers of conformal scalars, the isotropic cosmology with no scalar momenta is a special case, poised on the edge of a qualitatively different (and perhaps ill-defined) phase. On the positive side, this finding may turn out to be a selection principle, telling us that anisotropic or kinetic-dominated singularities should be excluded from the theory whereas isotropic universes with zero scalar momenta are allowed. If so, this would imply that black hole singularities, which locally resemble strongly anisotropic cosmological singularities, do not correspond to a bounce (contradicting the interpretation given by Ref.~\cite{Bars}, for example); there would be no ``born again" universe on the other side of the black hole singularity. On the negative side, one may wonder whether the inclusion of inhomogeneities could lead to problems even for the isotropic, nonkinetic cosmological bounce. We emphasize that, for $M=0$ or $M=1$, {\it any} amount of classical momentum in the zero modes of the anisotropy or scalar degrees of freedom takes the quantum mechanics of $a$ into the subcritical regime. Perhaps it is essential to work at $M>1$ for the theory to make sense. Clearly, we have only scratched the surface with this discussion, and there is a great deal to explore further. 

For the remainder of the paper, we {\it assume} that the quantum mechanics for $a$ makes sense. As explained in Ref.~\cite{letter}, this allows us to calculate the propagation of the Universe, and all inhomogeneous modes in it, by solving the theory on complex trajectories which bypass $a=0$ in the complex $a$ plane. Remarkably, as was also explained in Ref.~\cite{letter}, due to its scale-invariant property, the inverse square potential, if present, is actually {\it invisible} in our final results for ``in-out" amplitudes. 

\subsection{Feynman propagator}

Having defined positive- and negative-frequency modes by their asymptotics, given in Eq.~(\ref{hankelasymptotics}) (and {\it without} using any boundary condition at $a=0$), it is easy to obtain the Feynman propagator for the anisotropic case as a Green's function for the Wheeler-DeWitt equation, by using the Wronskian method as before. 

With the quantum Hamiltonian given by Eq.~(\ref{quantham}), the Feynman propagator satisfies
\bena
&&\left(-\Box+\xi\mathcal{R}+m^2\right)G(x,\lambda,m|x',\lambda',m')
\label{bianchiprop}
\\&=&-2\im m(-g)^{-\half}\delta^{M+1}(x-x')\delta^K(\lambda-\lambda')\delta(m-m')\nonumber
\eena
where we must introduce a factor $(-g)^{-\half}$ for the nontrivial metric determinant on superspace. Again switching to the scale factor coordinate $a$, Eq.~(\ref{bianchiprop}) is equivalent to
\bena
&&\left(\frac{\partial^2}{\partial a^2}+\frac{M+K}{a}\frac{\partial}{\partial a}-\frac{1}{a^2}\Delta_{H^M\times\bR^K}+\xi\mathcal{R}+m^2\right)G\nonumber
\\&=&-2\im m\frac{\delta(a-a')}{a^{M+K}}\frac{\delta^{M+K}(\nu-\nu',\lambda-\lambda')}{\sqrt{g_{H^M}}}\delta(m-m')\nonumber
\eena
with $G\equiv G(a,\nu,\lambda,m|a',\nu',\lambda',m')$, and the metric determinant on superspace is now made explicit. Again, we can now go to Fourier space on $H^M\times\bR^K$ introducing momenta $\zeta^i$ and $k^i$; the Feynman propagator in Fourier space satisfies
\bena
&&\left(\frac{\partial^2}{\partial a^2}+\frac{M+K}{a}\frac{\partial}{\partial a}-\frac{c}{a^2}+m^2\right)G(a,\zeta,k,m|a',\zeta,k,m')\nonumber
\\&=&-2\im m\,a^{-(M+K)}\delta(a-a')\delta(m-m')
\label{newwdweq}
\eena
with $c$ as in Eq.~(\ref{constant}). Since we have already identified the positive- and negative-frequency solutions (\ref{plusminus}) of the corresponding homogeneous equation, it is immediate to write down the solution to Eq.~(\ref{newwdweq}) with the correct boundary conditions,
\bena
G(a,m|a',m')&=&-2\im m\, a^{-(M+K)}\delta(m-m')\times\nonumber
\\&&W(\psi_-,\psi_+)^{-1}\left(\psi_-(a')\psi_+(a) \theta(a-a')\right.\nonumber
\\&&\left. +\psi_-(a)\psi_+(a') \theta(a'-a)\right)\,,
\eena
where the Wronskian is
\ben
W(\psi_-,\psi_+)=\frac{4a^{-(M+K)}}{\im\pi}
\een
and no longer constant in $a$, as is consistent with the appearance of a first derivative in Eq.~(\ref{wdwbianchi}). The Wronskian takes care of the factors of $a$ appearing in the elimination of the first derivative, Eq.~(\ref{psichi}), and cancels the determinant factor $a^{-(M+K)}$. The final result is
\bena
G(a,m|a',m')\,&&=\frac{\pi m}{2}\delta(m-m')(a\,a')^{-(M+K-1)/2}\times
\\&&\left(H_\nu^{(1)}(ma')H_\nu^{(2)}(ma)\theta(a-a')+(a\leftrightarrow a')\right)\nonumber
\eena
with $\nu\equiv \half\sqrt{1+4c'}$, which is consistent with the results of Ref.~\cite{letter} (with $K\equiv D-2$ as only the $D-2$ anisotropy degrees of freedom of a $D$-dimensional universe were considered there). One can check that in the absence of anisotropies or minimally coupled scalar fields, $K=0$, this result reduces to the expression obtained in Sec.~\ref{friedbounce}, {\it i.e.}, the propagator for a free massive particle in $(M+1)$-dimensional Minkowski spacetime. By our remarks below Eq.~(\ref{cpp}) the same should be true for $M=0$ or $M=1$ and general $K$, where the superspace metric is conformally flat.

\section{Perturbations}
\label{pertsec}

In this section, we extend our analysis to inhomogeneous cosmology, treated perturbatively at both linear and nonlinear order. We aim to solve the following problem: given an incoming state at large negative $a$ consisting of a flat, FRW, radiation-dominated classical background universe with perturbations in their local adiabatic vacuum state, what is the outgoing quantum state at large positive $a$, as defined by our analytic continuation prescription? This question can be answered, in the semiclassical limit, by using complex solutions of the classical Einstein-matter field equations. If one sends in any combination of linearized  positive- (negative-)frequency modes then, even after including the effects of nonlinearities in the field equations, it turns out that one finds only positive- (negative-)frequency linearized modes coming out. As we now explain,  this is sufficient to show, semiclassically, that the outgoing quantum state is also the local adiabatic vacuum. Hence, at a semiclassical level, there is no particle production across the bounce.  

Let us see this in detail. Consider a classical time-dependent background solution of the Einstein-matter equations.  If the matter is a perfect fluid, the only propagating degrees of freedom are scalar density perturbations and tensor gravitational wave modes. At the linearized level, we can decouple the modes by exploiting the homogeneity and isotropy of the background: for a flat background, every mode is a sum of plane waves $v(\eta,\vec{x}) =\sum_{\bf k} v_{\bf k} (\eta) e^{\im {\bf k}\cdot{\bf x}}$, with $v_{-\bf k} (\eta) =v_{\bf k} (\eta)^* $, with the coefficients decomposed into irreducible representations of the little group of rotations about ${\bf k}$. Now consider the action for the perturbations. At leading order, it is quadratic and it is diagonalized by the above mode decomposition. After a suitable time-dependent rescaling of the perturbations, the kinetic terms can always be brought to canonical form in which the action  reads [see, {\it e.g.}, Ref.~\cite{MFB}, page 269, Eq.~(10.59)]
\ben
{\cal S}^{(2)}=\sum_{{\bf k},a} \int d\eta \left(|\dot{v}_{\bf k}^a|^2-w_{k}^{2,a}(\eta) |v_{\bf k}^a|^2\right)\,,
\label{pert1}
\een
where the index $a$ labels the independent modes (here, scalar and tensor), and
\ben
w_{k}^{2,a}(\eta)= (k c_s^a)^{2} +m^{2,a}_{{\rm eff}}(\eta)
\een
where $c_s^a$ is the speed of sound, $1/\sqrt{3}$ for the scalar acoustic modes and unity for the tensor modes. In general, the time-dependent ``effective mass" introduces a nontrivial $\eta$-dependence. However, in our chosen background, the effective mass vanishes for both the scalar and tensor modes so $w_{k}^{a}=k c_s^a$ in both cases. 

We now make the assumption that the perturbations are well described by linear theory for wide intervals of conformal time $\eta$ well before and well after the bounce. As we see later, we cannot actually take the limit of infinite positive and negative conformal time because of the effect of nonlinearities in the fluid. Nevertheless, in the semiclassical approximation, and for modes whose wavelength is longer than the thermal wavelength of the fluid, the periods of incoming and outgoing conformal time during which linear theory remains valid are very large. We define our incoming and outgoing states during these intervals. 

When linear theory is valid, and when the frequencies $w_{k}^{a}(\eta)$ change adiabatically, $(d w_{k}^{a}/d\eta)/(w_k^a)^2 \ll 1$, the quantum states of the system are well described by those of a set of decoupled harmonic oscillators. Let us denote the corresponding real coordinates, {\it i.e.}, the real and imaginary parts of the $v_{\bf k}^a$, by the coordinates $q_m$, where the single index $m$ runs over all of the real, independent modes. Each of the coordinates $q_m$ contributes  an action ${\cal S}_m=\half\int d\eta (\dot{q}_m^2 -\omega_m(\eta)^2 q_m^2) $, and the adiabatic vacuum state is just the product of the corresponding harmonic oscillator ground states, 
\ben
\Psi_0(\eta,q)  = \prod_m (\omega_m /\hbar \pi)^{1/4} e^{-\omega_m q_m^2 /(2 \hbar)}.
\label{pert2}
\een
This state is uniquely defined by $a_m \Psi_0=0$ for all $m$, for the annihilation operator 
\ben
a_m\equiv \frac{1}{\sqrt{2 \omega_m \hbar}}\left(\hbar \frac{d}{d q_m} +\omega_m q_m\right) \,.
\een

Let us assume that the incoming state of the perturbations is $\Psi_{{\rm in}}(\eta',q)= \Psi_0(\eta',q)$ at some large negative $\eta'$, for which linear theory is valid. The quantum fluctuations in the fluid density may be shown to be small compared to the background density provided the wavelength of the modes is longer than the thermal wavelength, a condition which is in any case required in order for the fluid description to hold. The outgoing quantum state, at some large positive time $\eta$, is then given by propagating the incoming vacuum $\Psi_0$ to large positive times $\eta$, for which linear theory is once again valid, using the  path integral,
\ben
\Psi_{{\rm out}}(\eta,q) \approx{\cal N} \int \mathcal{D}q\; e^{\frac{\im}{\hbar} {\cal S}(q,\eta; q',\eta')} \prod_m dq'_m \Psi_{0}(\eta',q'), 
\label{pert3}
\een
where ${\cal S}(q,\eta; q',\eta')$ is the full, nonlinear Einstein-matter action taken with boundary conditions $q(\eta)=q$, $q(\eta')=q'$; $\mathcal{D}q$ indicates the complete path-integral measure and ${\cal N}$ is a normalization constant. We compute Eq.~(\ref{pert3}) in the semiclassical approximation, by finding the appropriate complex classical solution $q_m^c(\tilde{\eta}),$  $\eta'<\tilde{\eta}<\eta$, which is a stationary point of the combined exponent. Substituting Eq.~(\ref{pert2}) for $\Psi_{{\rm in}}$ and varying the exponent with respect to $q'_m$ yields, using the Hamilton-Jacobi relation, the initial condition for the classical solution $q^c$,
\ben
(\im p^c_m+\omega_m q^c_m)(\eta')=0,
\label{pert4}
\een
where $p^c_m=\dot{q}^c_m$ is the canonical momentum.  The initial condition (\ref{pert4}) specifies that $q^c_m$ is pure negative frequency at $\eta' $, a large negative time. The final boundary condition is just $q^c_m(\eta)=q_m$, where $\eta$ is a large positive time. We solve the classical Einstein-matter equations with these two boundary conditions in linear perturbation theory.  At linear order, the solution satisfying the boundary conditions is $q^c_m(\tilde{\eta})= q_m e^{\im k c_s (\tilde{\eta}-\eta)}$. Below, we give the complete solution for generic perturbation modes at linear and nonlinear order. We find that the solution is well described by linear perturbation theory at large negative and large positive times, with small nonlinear corrections, and that an incoming positive (negative) frequency mode evolves to an outgoing positive (negative) frequency mode which directly implies that the outgoing quantum state is the local adiabatic vacuum. To verify this, we need only apply the annihilation operators $a_m \propto \im p_m +\omega_m q_m = \hbar \frac{d}{d q_m} +\omega_m q_m$ to $\Psi_{{\rm out}}(\eta,q_m)$ as given in Eq.~(\ref{pert3}). Using the Hamilton-Jacobi equation, the result is proportional to $(\im p_m^c+\omega_m q_m^c)(\eta)$, which vanishes if the solution is pure negative frequency. Hence the incoming adiabatic vacuum evolves to the outgoing adiabatic vacuum, and there is no particle production across the bounce. 

\subsection{Basic setup and conventions}

We wish to study perturbations about a flat ($\kappa=0$) radiation-dominated FRW universe in a perturbation expansion. We shall go to nonlinear order but, for simplicity, 
restrict consideration to planar symmetry so that the metric depends only on conformal time $\eta$ and one spatial coordinate $x$, with two orthogonal spatial directions $(y,z)$. To keep the calculations manageable, we do not introduce conformally or minimally coupled scalar fields, so $M=K=0$. We work in Einstein gauge, {\it i.e.}, in the usual formulation of general relativity coupled to a radiation fluid.

The general form of the metric compatible with our assumed symmetry is
\begin{widetext}
\ben
ds^2=a^2(\eta)\begin{pmatrix} -1+\epsilon g_{\eta\eta}(\eta,x) & \epsilon g_{\eta x}(\eta,x) & & \\ \epsilon g_{\eta x}(\eta,x) & 1 + \epsilon g_{xx}(\eta,x) & & \\ & & 1+\epsilon g_{yy}(\eta,x) & \epsilon g_{yz}(\eta,x)\\ & &\epsilon g_{yz}(\eta,x)& 1+\epsilon g_{zz}(\eta,x)\end{pmatrix}\,.
\een
\end{widetext}

We can still apply coordinate transformations that leave this form of the metric invariant. A coordinate transformation $\eta=\tilde\eta+\epsilon\,g(x,\tilde\eta)$ changes the metric coefficients as
\bena
&&\delta g_{\eta\eta}=-2\left(\frac{\dot{a}}{a}g+\dot{g}\right)\,,\; \delta g_{\eta x} = -2 g'\,,\nonumber
\\&&\delta g_{xx}=\delta g_{yy}=\delta g_{zz}=2\frac{\dot{a}}{a}g
\eena
where here and in the remainder of this section ${}^{\cdot}$ is derivative with respect to $\eta$ and ${}'$ denotes derivative with respect to $x$. We use this freedom to eliminate $g_{\eta x}$ and introduce a different notation for the metric perturbation functions (note that in this section $\psi$ denotes a scalar metric perturbation, not a solution to the Wheeler-DeWitt equation as in earlier sections),
\bena
ds^2&=&a^2(\eta)\bigl[(-1+2\epsilon\phi(\eta,x))\,d\eta^2\nonumber
\\&&+(1+2\epsilon(\psi(\eta,x)+\gamma(\eta,x)))\,dx^2+\epsilon\, h^\times(\eta,x)\,dy\,dz\nonumber
\\&&+\left(1+\epsilon\left(2\psi(\eta,x)+\frac{h^T(\eta,x)}{2}\right)\right)\,dy^2\nonumber
\\&&+\left(1 + \epsilon\left(2\psi(\eta,x) - \frac{h^T(\eta,x)}{2}\right)\right)\,dz^2\bigr]\,.
\label{comov}
\eena
The form (\ref{comov}) is still left invariant by a transformation of the form
\bena
&&\eta=\tilde\eta+\epsilon\left(G(\tilde\eta)+\int^{\tilde{x}} dX\,\dot{f}(X,\tilde\eta)\right)\,,\nonumber
\\&&\quad x=\tilde{x}+\epsilon\,f(\tilde{x},\tilde\eta)\,,
\label{cotra2}
\eena
which we will use to simplify the matter variables. The energy-momentum tensor for radiation is
\ben
T_{\mu\nu}=\frac{4}{3}\rho u_{\mu}u_{\nu}+\frac{1}{3}\rho g_{\mu\nu}\,;\quad u^{\mu}u_{\mu}=-1\,.
\een
The density $\rho(\eta,x)$ and four-velocity $u^{\mu}(\eta,x)$ can also be written in terms of background and perturbation as
\bena
\rho(\eta,x)&=&\rho_0(\eta)(1+\epsilon\,\delta_r(\eta,x))\,,\nonumber
\\u^{\mu}(\eta,x)&=&\frac{1}{a(\eta)}(v^0(\eta,x),\epsilon v(\eta,x),0,0)\,.
\eena
The constraint $u^{\mu}u_{\mu}=-1$ can be solved for $v^0(\eta,x)$. Under a coordinate transformation (\ref{cotra2}), we have $\delta v=-\dot{f}$, so that we can set $v=0$ everywhere, {\it i.e.}, adopt a coordinate system in which the radiation is at rest everywhere ({\em comoving gauge}). The remaining gauge freedom is then under transformations
\bena
&&\eta=\tilde\eta+\epsilon\, G(\tilde\eta)\,,\quad x=\tilde{x}+\epsilon\,f(\tilde{x})\,,\label{gaugef}\\
&&y=\tilde{y}+\epsilon(\iota_1 \tilde{y}+\iota_2 \tilde{z}+\iota_3)\,,\quad z=\tilde{z}+\epsilon(\iota_4 \tilde{z}+\iota_5 \tilde{y}+ \iota_6)
\nonumber
\eena
where the $\iota_i$ are arbitrary constants and $G$ and $f$ are free functions. Under such a transformation, $\delta\phi=-\frac{\dot{a}}{a}G-\dot{G}$, $\delta\psi=\frac{\dot{a}}{a}G$, $\delta\gamma=f'$, $\delta h^T=2(a-b)$ and $\delta h^\times=2(c+d)$, and so functions of this form in the perturbations are to be considered pure gauge. We solve the Einstein equations in Fourier space, where the gauge freedom for the functions $\phi$, $\psi$, $h^T$ and $h^\times$ is somewhat hidden as it only becomes apparent for $k=0$.

We are left with five free functions for the metric ($\phi,\psi,\gamma,h^T,h^{\times}$) and the density perturbation $\delta_r$. As we will see, there are also six nontrivial Einstein equations relating these. To proceed, we assume that all of the perturbation functions can further be expanded as a power series in $\epsilon$,
\ben
\phi(\eta,x)=\sum_{n\ge 1}\epsilon^{n-1}\phi_n(\eta,x)\,,\quad{\rm etc.}
\een

The idea is now to solve the Einstein equations $G_{\mu\nu}=8\pi G T_{\mu\nu}$ order by order in $\epsilon$; the Einstein equations also imply energy-momentum conservation $\nabla_{\mu}{T^{\mu}}_{\nu}=0$ for the fluid. First, for the background (at order $\epsilon^0$) we have the equations
\ben
\dot\rho_0 + 4\frac{\dot{a}}{a}\rho_0 = 0\,,\quad \left(\frac{\dot{a}}{a^2}\right)^2 = \frac{8\pi G}{3}\rho\,.
\een
The first one tells us that $\rho_0\propto a^{-4}$ for some constant $M$; the Friedmann equation then gives the solution $a(\eta)\propto\eta$, the simplest example of a perfect bounce that we have already discussed in the introduction to this paper. It follows that $\frac{\dot{a}}{a}=\frac{1}{\eta}$, and that analytic continuation in the scale factor $a$ (as we have discussed in previous sections) is equivalent to analytic continuation in the conformal time coordinate $\eta$, which we use in this section.

At order $\epsilon^n$ in the perturbation expansion, the six nontrivial Einstein equations are 
\bena
\frac{3}{\eta^2}\delta_{r,n} - \frac{6}{\eta^2}\phi_n + 2 \psi_n'' - \frac{2}{\eta}\dot\gamma_n-\frac{6}{\eta}\dot\psi_n & = & J_{1,n}\,,
\label{einstnl1}
\\\frac{1}{\eta}\phi_n'+\dot\psi_n' & = & J_{2,n}\,,
\label{einstnl2}
\\\frac{1}{\eta^2}\delta_{r,n} - \frac{2}{\eta^2}\phi_n + \frac{2}{\eta}\dot\phi_n + \frac{4}{\eta}\dot\psi_n + 2 \ddot\psi_n & = & J_{3,n}\,,
\label{einstnl3}
\\\ddot{h}^{T}_n + \frac{2}{\eta}\dot{h}^T_n - (h^{T}_n)'' & = & J_{4,n}\,,
\label{einstnl4}
\\\ddot{h}^{\times}_n + \frac{2}{\eta}\dot{h}^{\times}_n - (h^{\times}_n)'' & = & J_{5,n}\,,
\label{einstnl5}
\\-\frac{1}{\eta^2}\delta_{r,n} + \frac{2}{\eta^2}\phi_n +\psi_n''-\phi_n''&&\nonumber
\\ - \frac{2}{\eta}\dot\gamma_n- \frac{2}{\eta}\dot\phi_n - \frac{4}{\eta}\dot\psi_n - \ddot\gamma_n - 2 \ddot\psi_n & = & J_{6,n}
\label{einstnl6}
\eena
for some ``source terms'' $J_{i,n}$ that are nonlinear combinations of the lower order perturbations. 

We first note that Eqs.~(\ref{einstnl4}) and (\ref{einstnl5}) that govern the tensor modes $h^{T}_n$ and $h^{\times}_n$ are already decoupled from the others. For the scalars, Eqs.~(\ref{einstnl1}) and (\ref{einstnl2}) can be solved for $\delta_r$ and $\phi$ directly. From Eq.~(\ref{einstnl2}) we get
\ben
\phi_n(\eta,x)=-\eta\dot\psi_n(\eta,x)+F_n(\eta)+\eta\int^x dx'\;J_{2,n}(\eta,x')
\label{phi}
\een
where $F_n(\eta)$ is a free function; then Eq.~(\ref{einstnl1}) implies that
\bena
\delta_{r,n}(\eta,x)&=&-\frac{2}{3}\eta^2\psi_n''+\frac{2}{3}\eta\dot\gamma_n+2F_n+\frac{\eta^2}{3}J_{1,n}\nonumber
\\&&+2\eta\int^x dx'\;J_{2,n}\,.
\label{delta}
\eena
Substituting these relations into Eqs.~(\ref{einstnl3}) and (\ref{einstnl6}) and taking linear combinations one obtains
\bena
\dot\gamma_n&=&\eta\psi_n''-6\dot\psi_n-3\dot{F}_n-\frac{\eta}{2}J_{1,n}-3\int^x dx'\,J_{2,n}\nonumber
\\&& -3\eta \int^x dx'\,\dot{J}_{2,n}+\frac{3\eta}{2}J_{3,n}
\label{gammeq}
\eena
and
\bena
\ddot\psi_n+\frac{2}{\eta}\dot\psi_n-\frac{1}{3}\psi_n''&=&-\frac{\dot{F}_n}{\eta}-\frac{\ddot{F}_n}{2}-\frac{J_{1,n}}{4}-\eta\frac{\dot{J}_{1,n}}{12}\label{psieq}
\\&&+\frac{\eta}{6}J_{2,n}'+\frac{11}{12}J_{3,n} + \frac{\eta}{4}\dot{J}_{3,n} +\frac{J_{6,n}}{6}\nonumber
\\&& -\int^x dx'\,\left(\frac{J_{2,n}}{\eta}+2\dot{J}_{2,n} + \frac{\eta}{2}\ddot{J}_{2,n}\right)\,.\nonumber
\eena
Equation (\ref{psieq}) can now be solved for $\psi_n$ using Green's functions; Eq.~(\ref{gammeq}) then gives $\gamma$ by a single integration over $\eta$, and from Eq.~(\ref{phi}) and Eq.~(\ref{delta}) one can obtain explicit expressions for $\delta_{r,n}$ and $\phi_n$ at each order. At each order in $\epsilon$, this provides an explicit algorithm for solving the system of Einstein equations (\ref{einstnl1})--(\ref{einstnl6}).

\subsection{Tensor perturbations}

Equations (\ref{einstnl4}) and (\ref{einstnl5}) are easily solved. First, consider the homogeneous equation solved by the first-order perturbation,
\ben
\ddot{h}^{T}_1 + \frac{2}{\eta}\dot{h}^{T}_1 - (h^{T}_1)''=0\,.
\label{homo}
\een
We can easily find the general solution in Fourier space, for $k\neq 0$,
\bena
h^{T}_1(\eta,x)&=&\int\frac{dk}{2\pi}e^{\im k x}h^{T}_1(\eta,k)\,,\nonumber
\\h^{T}_1(\eta,k)&=&b_1(k)\frac{e^{-\im k\eta}}{k\eta}+b_2(k)\frac{e^{\im k\eta}}{k\eta}\,.
\eena
For $k=0$, the two independent solutions are $h^T=\const$ and $h^T\sim 1/\eta$. We can write the general solution as
\ben
h^{T}_1=d_1+\frac{d_2}{k_0\eta}+\int\frac{dk}{2\pi}e^{\im k x}\left(b_1(k)\frac{e^{-\im k\eta}}{k\eta}+b_2(k)\frac{e^{\im k\eta}}{k\eta}\right)
\een
where $k_0$ is an arbitrary momentum scale to make $d_2$ dimensionless. $h^{\times}_1$ satisfies the same differential equation; its general solution is
\ben
h^{\times}_1=e_1+\frac{e_2}{k_0\eta}+\int\frac{dk}{2\pi}e^{\im k x}\left(c_1(k)\frac{e^{-\im k\eta}}{k\eta}+c_2(k)\frac{e^{\im k\eta}}{k\eta}\right)\,.
\een
For a real solution we need $b_1(k)=-b_1^*(-k)$, $b_2(k)=-b_2^*(-k)$ and similar for $c_1(k)$ and $c_2(k)$.

We recognize $d_1$ and $e_1$ as gauge modes corresponding to coordinate transformations (\ref{gaugef}), whereas $d_2$ and $e_2$ are physical $k=0$ modes. The free functions $b_1(k), b_2(k), c_1(k)$ and $c_2(k)$ are the physical gravitational degrees of freedom.

Now consider the general inhomogeneous equation,
\ben
\ddot{h}^{T}_n + \frac{2}{\eta}\dot{h}^{T}_n - (h^{T}_n)''=J_{4,n}\,.
\label{inhomo}
\een
Again, we go to Fourier space and first consider $k\neq 0$. We use the Wronskian method to determine a suitable Green's function; in contrast to the Green's function that appeared as a Feynman propagator in earlier sections, here the boundary conditions are that the higher order perturbations are set to 0 at some conformal time $\eta_0$ in the far past, so that only a linear (purely positive- or purely negative-frequency) mode is present. Given two independent solutions $h^T_1$ and $\tilde{h}^T_1$ to the homogeneous equation (\ref{homo}), the Green's function for these boundary conditions is
\ben
G(\eta,\eta')=\frac{h^T_1(\eta')\tilde{h}^T_1(\eta)-h^T_1(\eta)\tilde{h}^T_1(\eta')}{W(\eta')}
\een
for $\eta>\eta'>\eta_0$ where $\eta_0$ is the initial time at which only a linear perturbation is assumed to be present, and 0 otherwise. The Wronskian is $W(\eta')\equiv h^T_1 (\tilde{h}^T_1)'-(h^T_1)' \tilde{h}^T_1$ as before. Using this Green's function, we find that one particular solution to Eq.~(\ref{inhomo}) is
\ben
h^{T}_n(\eta,k)=\frac{1}{k\eta}\int^\eta d\eta'\;\eta'\;\sin(k(\eta-\eta'))\,J_{4,n}(\eta',k)\,,
\label{htfourier}
\een
while for $k=0$ we find
\ben
h^{T}_n(\eta,0)=\frac{1}{\eta}\int^\eta d\eta'\;\eta'\;(\eta-\eta')\,J_{4,n}(\eta',0)\,,
\label{htzero}
\een
which is just the limit $k\rightarrow 0$ of Eq.~(\ref{htfourier}). Expressions for $h^{\times}_n$ are analogous. In the integrals in Eqs.~(\ref{htfourier}) and (\ref{htzero}) as well as in the following, the initial time $\eta_0$ that should appear as the lower limit of integration is suppressed for simplicity; we neglect the $\eta_0$-dependent contributions as we are only interested in a particular solution.

Clearly, at each order $\epsilon^n$ one can also add a solution of the homogeneous equation to this solution for $h^T_n$. This can however be absorbed into the linear perturbation $h^T_1$. We hence set these arbitrary solutions to the homogeneous equations to 0 for $n\ge 2$.

\subsection{Scalar perturbations}

For the scalar perturbation functions $\phi,\psi,\gamma$ and $\delta_r$, we proceed analogously. For clarity, we first derive the general solutions for the first-order perturbations, for which there are no sources and the general solution is obtained straightforwardly. The equation for $\psi_{1}$ is Eq.~(\ref{psieq}) with the sources set to 0, {\it i.e.},
\ben
\ddot\psi_{1}+\frac{2}{\eta}\dot\psi_{1}-\frac{1}{3}\psi_{1}''=-\frac{\dot{F}_1}{\eta}-\frac{\ddot{F}_1}{2}
\een
where $F_1$ is a free function of $\eta$. Going into Fourier space, the general solution for $k\neq 0$, where $F_1$ does not contribute, is
\bena
\psi_1(\eta,x)&=&\int\frac{dk}{2\pi}e^{\im k x}\psi_1(\eta,k)\,,\nonumber
\\\psi_1(\eta,k)&=&a_1(k)\frac{e^{-\frac{\im}{\sqrt{3}}k\eta}}{k\eta}+a_2(k)\frac{e^{\frac{\im}{\sqrt{3}}k\eta}}{k\eta}\,.
\eena
The Fourier mode $k=0$ is a gauge mode [see the discussion below Eq.~(\ref{gaugef})], with general solution
\ben
\psi_1(\eta,0)=-\frac{c_1}{k_0\eta}+c_2-\frac{F_1(\eta)}{2}\,;
\een
since $F_1$ is arbitrary, we can set $c_1=c_2=0$ with no loss of generality. Putting this together, we have
\bena
\psi_1(\eta,x)&=&\int\frac{dk}{2\pi}e^{\im k x}\left(a_1(k)\frac{e^{-\frac{\im}{\sqrt{3}}k\eta}}{k\eta}+a_2(k)\frac{e^{\frac{\im}{\sqrt{3}}k\eta}}{k\eta}\right)\nonumber
\\&&-\frac{F_1(\eta)}{2}\,,
\eena
where for a real solution we need $a_1(k)=-a_1^*(-k)$ and $a_2(k)=-a_2^*(-k)$.

As said, from this expression we can determine the other scalar functions $\gamma,\phi$ and $\delta_r$. We find
\bena
\gamma_1(\eta,x)&=&a_3(x)+\int\frac{dk}{2\pi}e^{\im k x}\bigl[a_1(k)e^{-\frac{\im k\eta}{\sqrt{3}}}\left(-\frac{6}{k\eta}-\im\sqrt{3}\right)\nonumber
\\&&+a_2(k)e^{\frac{\im k\eta}{\sqrt{3}}}\left(-\frac{6}{k\eta}+\im\sqrt{3}\right)\bigr]\,,
\\\phi_1(\eta,x)&=&F_1(\eta)+\frac{\eta}{2} \dot{F}_1(\eta)+\int\frac{dk}{2\pi}e^{\im k x}\bigl[a_1(k)e^{-\frac{\im k\eta}{\sqrt{3}}}
\\&&\times\left(\frac{1}{k\eta}+\frac{\im}{\sqrt{3}}\right)+a_2(k)e^{\frac{\im k\eta}{\sqrt{3}}}\left(\frac{1}{k\eta}-\frac{\im}{\sqrt{3}}\right)\bigr]\,,\nonumber
\\\delta_{r,1}(\eta,x)&=&2F_1(\eta)+\int\frac{dk}{2\pi}e^{\im k x}\bigl[a_1(k)e^{-\frac{\im k\eta}{\sqrt{3}}}\left(\frac{4}{k\eta}+\frac{4\im}{\sqrt{3}}\right)\nonumber
\\&&+a_2(k)e^{\frac{\im k\eta}{\sqrt{3}}}\left(\frac{4}{k\eta}-\frac{4\im}{\sqrt{3}}\right)\bigr]\,,
\eena
where we recognize $a_3(x)$ and $F_1(\eta)$ as encoding the remaining gauge freedom in comoving gauge (\ref{gaugef}). $a_1(k)$ and $a_2(k)$ correspond to the scalar degrees of freedom of the radiation fluid.

In order to obtain the solutions for higher order perturbations, we derive the Green's function for the $\psi$ equation (\ref{psieq}), which has the general form 
\ben
\ddot\psi_n+\frac{2}{\eta}\dot\psi_n-\frac{1}{3}\psi_n''=\mathfrak{J}_n\,.
\een
Again, the boundary condition for the Green's function is to set the higher order perturbations to 0 at some initial conformal time $\eta_0$. We find, for $k\neq 0$,
\ben
\psi_n(\eta,k)=\frac{\sqrt{3}}{k\eta}\int^{\eta} d\eta'\;\eta'\;\sin\left(\frac{k(\eta-\eta')}{\sqrt{3}}\right)\,\mathfrak{J}_n(\eta',k)\,,
\label{psifourier}
\een
and for $k=0$ the same as for the tensors,
\ben
\psi_n(\eta,0)=\frac{1}{\eta}\int^\eta d\eta'\;\eta'\;(\eta-\eta')\,\mathfrak{J}_n(\eta',0)\,.
\label{psizero}
\een
Again, once the solution for $\psi_n$ is found, the other perturbation functions $\gamma_n$, $\phi_n$ and $\delta_{r,n}$ can be obtained easily.

From these expressions, we can now work out the nonlinear solution for all metric perturbation functions and the density perturbation order by order in $\epsilon$; all we need to do is to expand Einstein equations up to any given order to find the sources $J_{i,n}$ and then compute the integrals (\ref{htfourier}), (\ref{htzero}), (\ref{psifourier}) and (\ref{psizero}) to find the perturbations at the next order.

\subsection{Nonlinear positive-frequency modes}

We are now specifically interested in the nonlinear extension of linear positive-frequency modes at a given wave number $k_0$. (The following calculations and discussion can be extended to the case of an incoming negative-frequency mode by simply taking the complex conjugate of all expressions below, and replacing the lower- by the upper-half complex plane etc.) We choose the linear positive-frequency modes to be
\bena
\psi_1(\eta,x)&=&A\cos(k_0 x)\frac{e^{-\im k_0 \eta/\sqrt{3}}}{k_0 \eta}\,,\nonumber
\\h^T_1(\eta,x)&=&B\cos(k_0x)\frac{e^{-\im k_0 \eta}}{k_0 \eta}\,.
\eena
As seen before, the expression for $\psi_1$ then determines $\gamma_1, \phi_1$ and $\delta_{r,1}$. These scalar quantities are all gauge dependent but one may compute the gauge-invariant Newtonian potentials first introduced by Bardeen \cite{bardeen} and given by (in Fourier space for $k\neq 0$)
\ben
\Phi=-\phi+\frac{\ddot\gamma}{k^2}+\frac{\dot\gamma}{k^2 \eta}\,,\quad \Psi=-\psi-\frac{\dot\gamma}{k^2 \eta}\,.
\een
We find that for the linear perturbations $\Phi$ and $\Psi$ are equal and fall off as $1/k_0^2 \eta^2$ at large $|\eta|$,
\ben
\Phi_1(\eta,k_0)=\Psi_1(\eta,k_0)=-\frac{2\pi\,A\,e^{-\frac{\im}{\sqrt{3}}k_0 \eta}(3+\sqrt{3}\im\,k_0 \eta)}{k_0^3 \eta^3}\,.
\label{newt1}
\een

The explicit form of the sources at order $\epsilon^2$ is, in terms of the linear perturbations,
\begin{widetext}
\bena
J_{1,2} & = & \frac{12}{\eta^2}\phi^2 + \frac{3}{16}((h^T)')^2 + \frac{3}{16}((h^\times)')^2 + 2 \gamma'\psi' + 3(\psi')^2 +\frac{1}{4}h^T(h^T)''+\frac{1}{4}h^\times(h^\times)''+4\gamma\psi'' + 8\psi\psi'' - \frac{4}{\eta}\gamma\dot\gamma\nonumber
\\&& + \frac{4}{\eta}\phi\dot\gamma - \frac{4}{\eta}\psi\dot\gamma  -\frac{1}{2\eta}h^T\dot{h}^T-\frac{1}{16}(\dot{h}^T)^2 -\frac{1}{2\eta}h^\times\dot{h}^\times-\frac{1}{16}(\dot{h}^\times)^2 -\frac{4}{\eta}\gamma\dot\psi + \frac{12}{\eta}\phi\dot\psi-\frac{12}{\eta}\psi\dot\psi + 2\dot\gamma\dot\psi + 3\dot\psi^2\,,
\\J_{2,2} & = & \frac{2}{\eta}\gamma\phi' - \frac{2}{\eta}\phi\phi' + \frac{2}{\eta}\psi\phi' + \psi'\dot\gamma + \frac{1}{16}(h^T)'\dot{h}^T + \frac{1}{16}(h^\times)'\dot{h}^\times - \phi'\dot\psi + 2 \psi'\dot\psi + \frac{1}{8}h^T(\dot{h}^T)' + \frac{1}{8}h^\times(\dot{h}^\times)'\nonumber
\\&& + 2\gamma\dot\psi' + 4\psi\dot\psi'\,,
\\J_{3,2} & = & \frac{4}{\eta^2}\phi^2 - \frac{1}{16}((h^T)')^2 - \frac{1}{16}((h^\times)')^2 - 2\phi'\psi' + (\psi')^2 +\frac{1}{2\eta}h^T\dot{h}^T + \frac{3}{16}(\dot{h}^T)^2 + \frac{1}{2\eta}h^\times\dot{h}^\times + \frac{3}{16}(\dot{h}^\times)^2 - \frac{8}{\eta}\phi\dot\phi -\frac{8}{\eta}\phi\dot\psi\nonumber
\\&& + \frac{8}{\eta}\psi\dot\psi - 2\dot\phi\dot\psi + \dot\psi^2 + \frac{1}{4}h^T\ddot{h}^T+\frac{1}{4}h^\times\ddot{h}^\times -4\phi\ddot\psi + 4 \psi\ddot\psi\,,
\\J_{4,2} & = & -\gamma'(h^T)' - (h^T)'\phi' - 3 (h^T)'\psi' - 2 \gamma (h^T)'' - 4\psi(h^T)'' - 2 h^T\psi'' -\frac{4}{\eta}\phi\dot{h}^T + \frac{4}{\eta}\psi\dot{h}^T -\dot\gamma \dot{h}^T - \dot{h}^T\dot\phi + \frac{4}{\eta}h^T\dot\psi + \dot{h}^T \dot\psi\nonumber
\\&& - 2 \phi\ddot{h}^T + 2\psi\ddot{h}^T + 2 h^T\ddot\psi\,,
\\J_{5,2} & = & -\gamma'(h^\times)' - (h^\times)'\phi' - 3 (h^\times)'\psi'- 2 \gamma(h^\times)'' - 4\psi(h^\times)'' - 2h^\times\psi'' -\frac{4}{\eta}\phi\dot{h}^\times  + \frac{4}{\eta}\psi\dot{h}^\times -\dot\gamma \dot{h}^\times  - \dot{h}^\times\dot\phi  + \frac{4}{\eta}h^\times\dot\psi + \dot{h}^\times \dot\psi\nonumber
\\&& -2\phi\ddot{h}^\times + 2 \psi\ddot{h}^\times + 2 h^\times\ddot\psi \,,
\\J_{6,2} & = & -\frac{4}{\eta^2}\phi^2 + \frac{1}{16}((h^T)')^2  + \frac{1}{16}((h^\times)')^2 - \gamma'\phi' +(\phi')^2 +\gamma'\psi' + 2(\psi')^2 + \frac{1}{8}h^T(h^T)'' + \frac{1}{8}h^\times(h^\times)'' -2\gamma\phi'' + 2\phi\phi''\nonumber
\\&& - 2\psi\phi'' + 2\gamma\psi'' + 4 \psi\psi'' -\frac{4}{\eta}\gamma\dot\gamma + \frac{4}{\eta}\phi\dot\gamma -\frac{4}{\eta}\psi\dot\gamma -\dot\gamma^2 - \frac{1}{4\eta}h^T\dot{h}^T - \frac{1}{16}(\dot{h}^T)^2 - \frac{1}{4\eta}h^\times\dot{h}^\times - \frac{1}{16}(\dot{h}^\times)^2 + \frac{8}{\eta}\phi\dot\phi + \dot\gamma\dot\phi \nonumber
\\&& - \frac{4}{\eta}\gamma\dot\psi + \frac{8}{\eta}\phi\dot\psi - \frac{8}{\eta}\psi\dot\psi -\dot\gamma\dot\psi + 2\dot\phi\dot\psi - \dot\psi^2 -2\gamma\ddot\gamma + 2 \phi\ddot\gamma -2\psi\ddot\gamma - \frac{1}{8}h^T\ddot{h}^T - \frac{1}{8}h^\times\ddot{h}^\times - 2 \gamma\ddot\psi + 4 \phi\ddot\psi - 4 \psi\ddot\psi\,,
\eena
\end{widetext}
where we omit the subscripts ${}_1$ on the first-order perturbations on the right-hand side of these equations. For simplicity, we also set the second tensor mode $h^\times$ to 0 from now on (its dynamics are analogous to those of $h^T$).

We now compute the second-order perturbations from Eqs.~(\ref{htfourier}), (\ref{htzero}), (\ref{psifourier}) and (\ref{psizero}), using the sources computed from the chosen linear perturbations. For $h^T_2$, we find 
\bena
h^T_2&=&AB\left(\frac{3e^{-\im(1+\frac{1}{\sqrt{3}})k_0\eta}}{2k_0^2\eta^2}(1+\cos(2k_0x))\right.\nonumber
\\&&-\frac{\im(1+\sqrt{3})e^{-\im(1+\frac{1}{\sqrt{3}})k_0\eta}}{2k_0\eta}\cos(2k_0x)\nonumber
\\&&+\frac{\im(3+\sqrt{3})e^{-\im(1+\frac{1}{\sqrt{3}})k_0\eta}}{6k_0\eta}+\frac{2}{3}\Gamma\left(0,\im(1+\frac{1}{\sqrt{3}})k_0\eta\right)\nonumber
\\&&+\frac{13\im e^{-2\im k_0\eta}\,\Gamma(0,\im(-1+\frac{1}{\sqrt{3}})k_0\eta)}{6k_0\eta}\cos(2k_0x)\nonumber
\\&&\left.-\frac{13\im e^{2\im k_0\eta}\,\Gamma(0,\im(3+\frac{1}{\sqrt{3}})k_0\eta)}{6k_0\eta}\cos(2k_0x)\right)\,.\label{htexpr}
\eena
Here $\Gamma(0,z)$ are incomplete gamma functions. Their asymptotic expansion for large arguments is
\ben
\Gamma(0,z)\sim e^{-z}\left(\frac{1}{z}-\frac{1}{z^2}+O\left(\frac{1}{z^3}\right)\right)\,.
\een
Using this expansion, we see that as $k_0\eta\rightarrow\pm\infty$, $h^T_2$ has the asymptotic behavior
\bena
h^T_2 &&\sim AB\,e^{-\im(1+\frac{1}{\sqrt{3}})k_0 \eta} \Big(\im\frac{6+5\sqrt{3}-(27+16\sqrt{3})\cos(2k_0x)}{(21+11\sqrt{3}) k_0\eta}\nonumber
\\&&+\frac{23+4\sqrt{3}-(37+20\sqrt{3})\cos(2k_0x)}{2(5+2\sqrt{3})k_0^2 \eta^2}+O\left(\frac{1}{k_0^3 \eta^3}\right)\Big)\,.\nonumber
\eena
The asymptotic expansion shows in particular that all the terms in Eq.~(\ref{htexpr}) oscillate as $e^{-\im(1+\frac{1}{\sqrt{3}})k_0\eta}$ for large $k_0|\eta|$, and decay exponentially for large negative imaginary $\eta$.

We can obtain expressions for the scalar perturbations in exactly the same way. The expressions are similar to those for $h_2^T$ but involve more terms (15 in total), as there can be contributions of order $A^2$ and $B^2$, corresponding to two tensor modes or two scalar modes combining to give a scalar. Just as the second-order tensors, they contain incomplete gamma functions, but there is also a term involving a logarithm,
\ben
-A^2\,e^{-\frac{2}{\sqrt{3}}\im k_0 \eta}\frac{\im\sqrt{3}\,\log(k_0\eta)\,\cos(2k_0x)}{2k_0\eta}\,.
\een
These terms are potentially problematic when the perturbation functions are extended to the complex $\eta$ plane as the logarithms and incomplete gamma functions have branch cuts. However, all we require for positive-frequency modes is analyticity in the lower-half $\eta$ plane, where these modes extend to Euclidean, asymptotically decaying modes. This can be achieved by defining all the branch cuts to be along the positive imaginary axis. The analytic continuation of these modes that avoids the singularity at $\eta=0$ is then defined by choosing any contour that remains in the lower-half complex $\eta$ plane.

Asymptotically, we find that at large $k_0|\eta|$,
\bena
\psi_2 &\sim& -e^{-2\im k_0 \eta}\,\frac{7\im\, B^2}{128 k_0 \eta} + e^{-\frac{2}{\sqrt{3}}\im k_0 \eta}A^2\left(\frac{1}{12}+\frac{1}{6}\cos(2k_0 x)\right.\nonumber
\\&&\left.-\im\frac{2+\cos(2k_0x)(1+12\log(k_0 \eta))}{8\sqrt{3}k_0 \eta}\right)+O\left(\frac{1}{k_0^2 \eta^2}\right)\,,\nonumber
\eena
and one can check that all terms, including all subleading ones, oscillate at positive frequencies asymptotically (either at $\omega=2k_0$ or at $\omega=\frac{2}{\sqrt{3}}k_0$). The nonlinear modes again decay exponentially as $k_0\eta\rightarrow -\im\infty$, and indeed define nonlinear positive-frequency modes. From the general structure of the equations (\ref{einstnl1})--(\ref{einstnl6}), one can see that the same property should hold to all higher nonlinear orders: the source terms, being nonlinear in lower order perturbations, always decay exponentially sufficiently fast in imaginary time that integration with a Green's function that has an exponentially growing and an exponentially decaying part, as in Eq.~(\ref{psifourier}), gives again an exponentially decaying next-order perturbation. The method we have described then allows a general definition of positive-frequency modes in the complex $\eta$ plane, to all orders in perturbation theory.

The other perturbations are determined by Eqs.~(\ref{phi})--(\ref{gammeq}). For completeness, we give their asymptotic expressions for large $k_0|\eta|$,
\bena
\gamma_2 &\sim& A^2\,e^{-\frac{2}{\sqrt{3}}\im k_0 \eta}\left(-\frac{\im\cos(2k_0x)k_0\eta}{\sqrt{3}}\right.
\\&&\left.-1-\cos(2k_0x)\left(\frac{5}{4}+3\log(k_0 \eta)\right)\right)+O\left(\frac{1}{k_0 \eta}\right)\,,\nonumber
\\\delta_{r,2} &\sim& \frac{1}{3}A^2\,e^{-\frac{2}{\sqrt{3}}\im k_0 \eta}\Big(\frac{4\im\cos(2k_0x)k_0\eta}{\sqrt{3}}
\\&&-7-\cos(2k_0x)(7-12\log(k_0 \eta))\Big)+O\left(\frac{1}{k_0 \eta}\right)\,,\nonumber
\\\phi_2 &\sim& e^{-2\im k_0 \eta}\,\frac{7 B^2}{64}  + A^2\,e^{-\frac{2}{\sqrt{3}}\im k_0 \eta}\left(\frac{\im k_0 \eta(1+2\cos(2k_0 x))}{6\sqrt{3}}\right.\nonumber
\\&&\left.+\frac{1}{3}-\cos(2k_0 x)\left(\frac{1}{12}-\log(k_0\eta)\right)\right)+O\left(\frac{1}{k_0 \eta}\right)\,.\nonumber
\eena
To verify the validity of our solution method, we have checked explicitly that the second-order perturbations solve Einstein's equations up to order $\epsilon^2$.

We see that none of the scalar perturbation functions decay at real infinity $k_0|\eta|\rightarrow\infty$, and some even blow up, indicating a breakdown of perturbation theory at large times. Again, to get gauge-invariant statements about this behavior, we can compute the Newtonian potentials, and find that they fall off as $1/k_0 \eta$,
\ben
\Phi_2(\eta,2k_0) \sim -e^{-\frac{2}{\sqrt{3}}\im k_0 \eta}\frac{\im\pi\sqrt{3}a^2}{4k_0 \eta}+O\left(\frac{1}{k_0^2 \eta^2}\right)
\een
with similar behavior for $\Psi$. This compares with $\Phi_1\sim O(1/k_0^2\eta^2)$ in Eq.~(\ref{newt1}), which still indicates that the perturbation expansion breaks down when $\epsilon k_0|\eta|\sim 1$. This physical behavior is due to the nonlinear evolution in the fluid, as shown by analytical and numerical studies in Ref.~\cite{fluidpaper}. When we go down the imaginary axis, {\it i.e.} for $\eta=-\im\tau$ with $\tau\rightarrow\infty$, all perturbation functions fall off exponentially, with exponential terms of the form $e^{-\omega\tau}$ dominating any polynomially growing terms. As we have argued, this behavior persists for higher orders in the $\epsilon$ expansion, and defines these modes by regularity for large negative imaginary $\eta$; the blowup of scalar perturbations along the real axis due to nonlinearities in the fluid does not prevent us from defining nonlinear asymptotic positive-frequency modes.

\subsection{Summary}

We have given an algorithm for solving the Einstein-matter equations order by order in perturbation theory, and exhibited explicit results at second order that show in detail how the positive-frequency incoming modes match only to positive-frequency outgoing modes, and similarly for negative-frequency modes (where our results trivially extend by taking complex conjugates). We have argued that this behavior should extend to all orders in perturbation theory, as the nonlinear extension of linear positive-frequency modes leads to perturbation functions that decay exponentially for large negative imaginary times, and branch cuts can be restricted to the positive half-plane for positive-frequency solutions, so that the nonlinear metric perturbation satisfies a nonlinear notion of positive frequency. We identified some subtleties, namely that the perturbation expansion fails at late times $k_0|\eta| \sim \frac{1}{\epsilon}$, where $\epsilon$ is the perturbation amplitude, meaning that one has to restrict attention to an annulus in the complex plane, $\epsilon<k_0|\eta|<\frac{1}{\epsilon}$, in which the $\epsilon$ expansion can be trusted and nonlinearities are not yet dominant \cite{letter}. 

\section{Conclusions}
\label{conclsec}

This paper represents a detailed study of a very simple cosmological model, based on the principle of conformal symmetry for matter and gravity and the observed fact that the early Universe was dominated by radiation. Classical cosmological solutions of this model describe a bounce, with a big bang/big crunch singularity, but the singularity can be avoided by going into the complex plane. While this ``singularity avoidance'' seems {\em ad hoc} in classical gravity, we have shown its meaning in the quantum theory where, similar to quantum tunneling, the complexified solutions represent legitimate saddle points to the path integral. The picture that emerges for quantum cosmology is based on modes that are asymptotically purely positive frequency at early and late times when the Universe is large and classical, corresponding to a positive expansion rate of the Universe, as we observe. We have shown that the addition of a positive radiation density makes a crucial difference, as it leads to classical solutions which connect asymptotic contracting and expanding Lorentzian regions, and which are represented by the positive-frequency modes defined by the Feynman propagator. We do not impose any boundary conditions for the wave function at $a=0$, and accept that some modes may even diverge there: all that is required is a consistent evolution from an asymptotic contracting to an asymptotic expanding universe, through or around the bounce, as this allows a calculation of transition amplitudes and hence, ultimately, predictions for the transition of a given state in the contracting phase to a state in the expanding phase. This formalism appears much more natural than an imposition of a boundary condition at $a=0$, where quantum effects are large and where classical notions of singularity avoidance  may cease to have any relevance. In practical terms, the fact that our wave functions and propagators admit a semiclassical WKB description in which high-curvature regions near $a=0$ can be avoided gives hope that a semiclassical approach to the quantum cosmology of bouncing scenarios can be used for predictions, even in the absence of a complete theory of quantum gravity.

Some features we are exploiting are clearly restricted to homogeneous cosmological models such as the FRW and Bianchi I universes we have studied explicitly. It is therefore vital to check that the formalism can be extended consistently to generic perturbations around homogeneity, and ultimately to fully nonlinear solutions of GR. We have developed a systematic perturbative treatment that shows how this question can be attacked, at linear and nonlinear order, and given evidence for a consistent nonlinear extension of positive-frequency modes to the complex $a$ plane. Again, one is interested in the transition of incoming asymptotic positive-frequency modes to outgoing modes which are, in general, a mixture of positive and negative frequency and which signal particle production (and potential divergencies) at the bounce. We have shown that an incoming positive-frequency mode can be continued around the singularity, and unambiguously matches to an outgoing positive-frequency solution. So the incoming adiabatic vacuum state is stable across the bounce and no divergencies arise. Our calculations have been limited to pure radiation and planar symmetry, and one focus of future work will be to extend these results to more general cases. The present results already indicate that a consistent semiclassical picture exists for nonlinear perturbations of cosmological models, and that this picture can be used for calculations of the cosmological phenomenology of bounce scenarios of the type we consider.

Thus, our results show how classical singularities do not necessarily prevent a consistent quantum description of bouncing cosmologies. The inclusion of quantum effects into the big bounce seems a natural and simple alternative to the development of more complicated bounce scenarios~\cite{matterbounce,modgravbounce,qgbounce,lehnerslatest}. 

There are many avenues for further exploration. In Sec.~\ref{anisosec}, we began to explore the quantum theory on the real $a$-axis around $a=0$. In some cases, it may be that the attractive inverse square potential in the Wheeler-DeWitt operator may lead the quantum theory to fail when further (inhomogeneous) degrees of freedom are included, but in others the quantum theory seems to be healthy.  The quantum dynamics of more general Bianchi models also deserve to be  understood; for these, the invisibility of the singularity that we have observed for Bianchi I will presumably be replaced by a nontrivial scattering matrix between in and out asymptotic states. The pathologies we have identified in the Feynman propagator for curved FRW universes should be revisited with the inclusion of a positive cosmological constant. More basic conceptual questions concerning the interpretation of the propagator and the determination of probabilities need to be investigated. Ultimately, we need to find a compelling measure on the space of quantum universes.  There are hints that the present flat, isotropic universe lies on a critical boundary in the quantum theory, and these may point to novel resolutions of the
classic flatness and isotropy puzzles. 

\acknowledgments

We thank A.~Ashtekar, I.~Bars, J.~Halliwell, J.-L.~Lehners, L.~Smolin, P.~J.~Steinhardt and especially J.~Feldbrugge for helpful discussions, and L.~Sberna for comments on the manuscript. One of us (N.T.) thanks  E.~Witten for encouraging remarks. 

This research was supported in part by Perimeter Institute for Theoretical Physics, in particular through the Mike and Ophelia Lazaridis Niels Bohr Chair. Research at Perimeter Institute is supported by the Government of Canada through the Department of Innovation, Science and Economic Development Canada  and by the Province of Ontario through the Ministry of Research, Innovation and Science. The work of S.G. was supported in part by the People Programme (Marie Curie Actions) of the European Union's Seventh Framework Programme (FP7/2007-2013) under REA grant agreement n$^{{\rm o}}$ 622339. 

\appendix

\section{Massive Relativistic Propagator} 

In this appendix, we calculate the massive relativistic propagator given in Eq.~(\ref{e2}) exactly. First, we note that 
\ben
G(x| x') =\im\int_0^{\infty}  d\tau \left(\frac{m}{2\pi\im\tau}\right)^{\frac{M+1}{2}}e^{-\im \frac{m}{2}\left(\frac{\sigma}{\tau}+\tau\right) }\,,\nonumber
\een
is a convergent integral when $\sigma=-(x-x')^2$ is positive. The $\tau$ integral may be taken along the positive real axis $0<\tau<\infty$. Next, we set $\tau=\sqrt{\sigma} e^u$, with $-\infty<u<\infty$, so that 
\bena
G(x| x')&=& \im\int_{-\infty}^{\infty} du (\sqrt{\sigma})^{\frac{1-M}{2}} \left(\frac{m}{2 \pi \im} \right)^{\frac{M+1}{2}} e^{ -\im m \sqrt{\sigma} {\rm cosh} u - \frac{M-1}{2} u} \nonumber\\
&=& \half (-\im m)^M (2 \pi m \sqrt{\sigma})^{\frac{1-M}{2}} H_{\frac{M-1}{2}}^{(2)}(m \sqrt{\sigma})\,,
\eena
where we have used the standard integral representation of the Hankel function of the second kind,
\ben
H_\nu^{(2)}(z)=\frac{\im^{\nu+1}}{\pi}\int\limits_{-\infty}^{\infty}du\;e^{-\im z \cosh u-\nu u}\,,
\een
and for positive real argument the function is defined as the boundary value of a function in the lower-half complex $z$ plane where the integral converges.

Following the discussion given in Sec.~\ref{feynman}, the result is then continued to negative values of $\sigma$ by analytic continuation through the lower-half complex $\sigma$ plane.

\end{document}